\documentstyle[epsf]{article}

\begin{document}
\begin{flushright} {\bf CU-TP-876}
\end{flushright}
\vskip 20pt

\begin{center}
{\Large\bf Gluon production in current-nucleus and nucleon-nucleus
collisions in a quasi-classical approximation\footnotetext{*This work
is supported in part by the Department of Energy under GRANT
DE-FG02-94ER-40819.}}

\vskip 20pt

Yuri V. Kovchegov*  \\
and\\
A.H. Mueller*
\\~\\

{\it Physics Department, Columbia University\\ New York, New York
10027}
\end{center}

\vskip 24pt
\begin{abstract}
 We calculate gluon production in deep inelastic scattering of the
current $j= - {1\over 4} F_{\mu\nu}^a F_{\mu\nu}^a$ off a large
nucleus and in nucleon-nucleus collisions.  In a covariant gauge
calculation the transverse momentum spectrum of the gluon is
determined by the final state interactions of the gluon with the
nucleons in the nucleus. In a light-cone gauge calculation final state
interactions are absent and these effects come from the light-cone
wavefunction of the nucleus.  We work in an approximation which
neglects QCD evolution of gluons in the nucleon, that is in a
quasi-classical approximation.
\end{abstract}

\section{Introduction}

A few years ago L. McLerran and R. Venugopalan\cite{LMc} introduced an
interesting program of studying the small x gluon distribution, and
gluon production, involving a very large nucleus.  The essential idea
is to use the valence quarks of the nucleus as a source for a
light-cone gauge calculation, involving only tree graphs, of the
Weizs\"acker-Williams(WW) field of the nucleus.  The higher momentum
components of this WW field, interpreted in terms of gluon quanta,
would then be included with the valence quarks to give a new source
which could be used to calculate another momentum layer of WW quanta,
etc.  This program is being actively pursued\cite{JJa,JJM}.  The
calculation of the WW field of the valence quarks of a large nucleus
is a nontrivial problem because of the nonlinearity due to the strong
source coming from the many valence quarks of the large nucleus.  It
was calculated independently in \cite{YuV} and \cite{JJK}.

It was observed in this (quasi-classical) WW approximation that the
gluon distribution for a large nucleus is strongly modified from what
one would expect from adding the distributions of the individual
nucleons incoherently.  An apparent saturation of the gluon
distribution sets in at moderate to low gluon transverse momenta while
the high transverse momentum part of the WW single particle gluon
distribution remains additive.  However, some time ago\cite{AHM} it
was observed that there are no shadowing corrections, in this
$equivalent\ gluon$ approximation. It is one of the purposes of this
paper to show the compatibility of the $no\ shadowing$ result with the
result of Ref.5 and to interpret that latter result.  To do this we
introduce a {\em current} $j(x) = - {1\over 4} F_{\mu\nu}^a(x)
F_{\mu\nu}^a(x)$ and calculate deep inelastic scattering of this
current off a large nucleus in an approximation where the nucleons in
the nucleus have no QCD evolution included in their gluon
distributions.  We suppose $\alpha \ll 1$ but the parameter
$\alpha^2A^{1/3}$, with \ A\ the atomic number, will generally be
taken to be large.  We do our calculation first in covariant gauge and
then in an appropriate light-cone gauge in order to interpret the
result of Ref.5.

It is convenient to consider two distinct regimes of gluon production
in the deep inelastic scattering of the current\ j\ off a large
nucleus.  If $Q^2 = - q_\mu q_\mu$ is the virtuality of the current
and if ${\sqrt{<\ell_\perp^2>}}$ is the typical transverse momentum
that a high energy gluon obtains by multiple scattering with the
nucleons as it passes through the nucleus\cite{GTB}, then these two
regimes are characterized by $Q^2 \gg <\ell_\perp^2>$ and by
${Q^2<}\ \ <\ell_\perp^2>.$ We begin by describing our results when
$Q^2 \gg <\ell_\perp^2>.$

In small\ x\ deep inelastic scattering, and in covariant gauge, the
current \ j\ produces a gluon locally in the nucleus\cite{AHM}.  This
gluon then multiply scatters, both elastically and inelastically, as
it passes through the nucleus\cite{RJG,VNG,MLJ}.  The equation for the
transverse momentum distribution of the produced gluon is given in (6)
below.  The solution, given in (18) and (21), is exactly the one found
in \cite{JJK} for the gluon correlation function.  This distribution
does not reflect shadowing, but rather the probability conserving
final state interactions which modify the transverse momentum
distribution of the produced gluon but not the cross section for its
production.  In an $A_+=0$ light-cone gauge calculation, in a frame
where the nucleus is right-moving and the current left-moving, we find
a complete absence of final state interactions and we give a physical
picture of why this happens.  Here the effect of the final state
interaction which we found in covariant gauge is encoded in the
light-cone wavefunction of the nucleus.  It is quite remarkable that
the distribution of gluons in the WW wavefunction gives the correct
transverse momentum distribution of produced gluons without any final
state interactions whatever. The traditional gluon distribution
$xG_A(x,Q^2)$ can be obtained by integrating the produced gluon
distribution over all transverse momenta obeying $\ell_\perp^2 \leq
Q^2.$ Because $Q^2 \gg \ell_\perp^2,$ in the logarithmic
approximation, one can view the current \ j\ as measuring the gluon
distribution in the nuclear wavefunction.

When ${Q^2<}\ <\ell_\perp^2>$ the covariant gauge calculation of the
produced gluon spectrum does not change in form from that when $Q^2
\gg <\ell_\perp^2>$ since the large $<\ell_\perp^2>$ comes from final
state interactions which appear after the production of the gluon
occurs.  The light-cone gauge calculation now appears more
complicated.  The three and four gluon terms in \ j\ become important
and final state interactions are not negligible.  The current is no
longer sufficiently pointlike to be viewed as measuring the partonic
gluon distribution of the nucleus.  If we construct $xG_A(x,Q^2)$ from
its moments, $defined$ in terms of nuclear matrix elements of the
normal local gluon operators, with the counter terms taken to be the
same as for nucleon matrix elements, then $xG_A(x,Q^2)$ is additive
$xG_A(x,Q^2) = A xG(x,Q^2).$ It appears that this object has no direct
connection with partonic gluons in the light-cone wavefunction of the
nucleus when ${Q^2<}\ <\ell_\perp^2>.$ It is perhaps surprising that
the gluon distribution in the light-cone wavefunction is still the
same as the produced gluon spectrum.  Finally, while the gluon
distribution in the light-cone wavefunction is very different from the
incoherent nucleon light-cone distributions when ${\ell_\perp^2<}\
<\ell_\perp^2>$ the nuclear modifications are simply probability
conserving redistributions in phase space and so are unrelated to
nuclear shadowing.

In Sec.3 we calculate gluon production in a nucleon-nucleus collision.
In a calculation equivalent to a covariant gauge calculation we view
the gluon as being produced in one of two ways.  (i) The gluon may be
present in the wavefunction of the hadron as the hadron reaches the
front face of the nucleus.  In that case interactions with nucleons in
the nucleus free the gluon and multiple scattering in the nucleus
broadens its transverse momentum spectrum.  (ii) The gluon may be
radiated in the final state from fragments of the incident hadron.  In
that case the transverse momentum spectrum of the gluon is not
affected by the nucleus. Interference terms between initial and final
state emission are small.  In an $A_+=0$ light-cone gauge calculation
of gluon production from the collision of a right-moving nucleus with
a left-moving nucleon no final state interactions appear.  If
$A_\mu^\perp$ is the field of the nucleus while $A_\mu^\prime$ is the
field of the nucleon gauge rotated\cite{YK} by those nucleons in the
nucleus having the same impact parameter then the gluon production
cross section is obtained by coupling $A_\mu^\perp$ and $A_\mu^\prime$
to the QCD radiation field in a way determined by the QCD Lagrangian.
A very simple expression emerges for the cross section as given by
(73)-(75).

\section {The gluon distribution and gluon production in deep inelastic
scattering off a large nucleus}

In this section, we consider deep inelastic scattering off a large
nucleus where the (gauge invariant) current initiating the scattering
is taken to be $j(x) = - {1\over 4} F_{\mu\nu}^i F_{\mu\nu}^i$ with
$F_{\mu\nu}^i$ the usual QCD field strength tensor.  We choose this
current because it furnishes the most direct way to measure the gluon
distribution and to produce a gluon and follow its interactions with
the nucleons of the nucleus as it passes into the final state.  We
neglect QCD evolution in the interaction of\ j\ with the individual
nucleons in the nucleus.  Our calculation of the gluon distribution is
closely related to that of Ref.5 with which we are in agreement.  Our
motivation for neglecting QCD evolution, and we shall explain below in
what sense we neglect it, is the same as in Ref.1.  We are interested
in a model to illustrate particular small\ x\ and nuclear effects.  We
shall later comment on our expectations in a realistic nucleus.

There is a surprisingly different picture of gluon production in
covariant gauge as compared to light-cone gauge. In covariant gauge,
and in the rest systems of the nucleus, the current\ j\ produces a
gluon off a nucleon in the nucleus in a reasonably local way.  This
gluon then has final state interactions with nucleons along its path
as it passes through the nucleus\cite{MLJ,RBY}.  The final state
interactions change the transverse momentum distribution of the
produced gluon, but they do not modify the production rate.  Thus the
gluon distribution, and the structure function, of the nucleus is
additive in the atomic number and there is no shadowing.  In
light-cone gauge, at least in a particular light-cone gauge, there
are no final state interactions whatever.  The transverse momentum
distribution of the produced gluon is already encoded in the
wavefunction of the nucleus. After deriving this result we shall give
a simple physical picture why final state interactions are not present
in our light-cone gauge.

\subsection{Covariant gauge calculation of gluon production of the nuclear
gluon distribution}

We begin with the scattering of $j = - {1\over 4} F_{\mu\nu}^2$ off a
single nucleon as illustrated in Fig.1.  Define a structure function

$$W = (2\pi)^32 E_p \int d^4x\ e^{iqx}(p\vert j(x) j(0)\vert
p)\eqno(1)
$$
\noindent in analogy with deep inelastic lepton nucleon scattering.
Then, from the graph in Fig.1 it is straightforward to show

$$
W(s,Q^2) = {s\over 4} W_0\ell n\ s/\mu^2\eqno(2a)
$$

\noindent which can be factorized as

$$W(s,Q^2) = {s\over 4} xG(x,Q^2) + {s\over 4} W_0 \ell n\ 1/x\eqno(2b)$$

\noindent with\ x\ the usual Bjorken variable, $s=(p + q)^2$ and $W_0$
a constant.  (The Feynman rules for coupling \ j\ to two gluons are
given in Ref.6.)  The first term on the right-hand side of (2b)
corresponds to gluon operators in the operator product expansion while
the second term corresponds to quark operators.  As discussed in some
detail in Ref.6 the nucleons are additive in our approximation so that

$$W_A = AW\eqno(3)$$

\noindent and

$$xG_A = A x G\eqno(4)$$

\noindent Where $W_A$ and $G_A$ are the structure function and the
gluon distribution for the nucleus.  Indeed, additivity is closely
related to our treatment of the gluon-hadron amplitude, the lower {\em
blob} in Fig.1.  We suppose that there are no $\ell n\ k^2/\wedge^2$
factors in that amplitude and we also assume that the gluon-nucleon
amplitude vanishes as $k\cdot p$ becomes large.  That is, we suppose
that high mass intermediate states are not present in the
gluon-nucleon scattering amplitude.  These $assumptions$ limit the
longitudinal coherence length of the forward current-nucleon
scattering amplitude to be not significantly larger than the nucleon
size, which means that the interaction is local in the nucleus.

Now let us follow a gluon, produced by an interaction of \ j\ with a
nucleon in the nucleus, as it scatters with other nucleons in the
nucleus as illustrated Fig.2 where we have shown examples of both
inelastic and elastic scattering in the nucleus.  It is convenient to
define a probability distribution for the gluon to have a transverse
momentum ${\underline \ell}$ as it passes through the
nucleus\cite{RBY}.  If the gluon is produced with a transverse
momentum ${\underline \ell}_0$ at a longitudinal coordinate $z_0$ and
at an impact parameter ${\underline b}$ with respect to the center of
the nucleus, let $f({\underline b},z_0,{\underline \ell}_0, z,
{\underline \ell})$ be the probability distribution for the gluon to
have transverse momentum ${\underline \ell}$ at a longitudinal
position z.  Of course

$$\int d^2\ell f({\underline b}, z_0, {\underline \ell}_0, z,
{\underline \ell}) = 1\eqno(5)$$

\noindent while the number density of produced gluons in the nucleus
is

$${d N(\ell)\over d^2\ell} = \int d^2\ell_0 {dN_0\over
d^2\ell_0}d^2bdz_0 \rho({\underline b},z_0) f({\underline
b},z_0,{\underline \ell}_0,z,{\underline \ell})\vert_{z={\sqrt{R^2-
b^2}}}\eqno(6)$$

\noindent where the integration $d^2bdz_0$ goes over the volume of the
nucleus, $\rho$ is the nuclear density normalized to the atomic
number\ A\ by

$$\int d^2bdz_0\rho({\underline b}, z_0) = A, \eqno(7)$$

\noindent and we take the nuclear density to be uniform throughout the
nucleus for simplicity.  The initial distribution resulting from an
interaction of the current \ j\ with a nucleon in the nucleus is
normalized to be given by the unintegrated gluon distribution of the
nucleon

$${dN_0({\underline \ell})\over d^2\ell} = {1\over \pi} {dN_0\over
d\ell^2} = {1\over \pi} {\partial\over \partial
\ell^2}[xG(x,\ell^2)]\eqno(8)$$

\noindent where, in our approximation, there is no x-dependence in
$xG(x,\ell^2)$ and there is no $\ell^2-$dependence in
$\ell^2{\partial\over \partial\ell^2} x G(x,\ell^2)$ for large
$\ell^2.$

Now \ f\ obeys the equation (See Eqs.2.7 and 2.8 of Ref.12. Appendix
A of that reference gives a derivation.)

$${\partial\over \partial z} f(z,{\underline \ell}) =\ - {1\over
\lambda} f(z,{\underline \ell})+ \rho\sigma \int d^2\ell^\prime
V({\underline \ell}^\prime) f(z,{\underline \ell}-{\underline
\ell}^\prime)\eqno(9)$$

\noindent with

$$f(z_0,{\underline \ell}) = \delta({\underline \ell}-{\underline
\ell}_0)\eqno(10)$$

\noindent and where

$$V({\underline \ell}) = {1\over \sigma} {d\sigma\over d^2
\ell}\eqno(11)$$

\noindent is the normalized gluon-nucleon scattering amplitude with
$\ell$ the momentum transfer and where

$$\lambda = [\rho\sigma]^{-1},\eqno(12)$$

\noindent is the mean free path of gluons in the nuclear matter.  For
notational simplicity we suppress the ${\underline b}, z_0,{\underline
\ell}_0$ dependence of f. The first (inelastic) interaction of the
produced gluon shown in Fig.2 is contained in the second term in (9)
while the second (elastic) interaction comes from the first term in
(9). We work in an approximation where the gluon-nucleon cross section
is energy independent, at high energy, and where the gluon's helicity
is conserved during the scatterings in the nucleus.

Eq.9 is solved by going to a transverse coordinate representation

$$\tilde{f}(z,{\underline x}) = \int d^2 \ell e^{-i({\underline
\ell}-{\underline \ell}_0) \cdot {\underline x}}f(z,{\underline
\ell}).\eqno(13)$$

\noindent As usual in the high energy limit the transverse coordinate
of the gluon will not change as the gluon passes through the nucleus,
and thus, using (13) in (9) gives

$${\partial\over \partial z} \tilde{f}(z,{\underline x}) = - {1\over
4\lambda} {\underline x}^2 \tilde{v}({\underline
x})\tilde{f}(z,{\underline x})\eqno(14)$$

\noindent with

$$\tilde{f}(z_0,{\underline x}) = 1\eqno(15)$$

\noindent and where

$$\tilde{v}({\underline x}) = {4\over {{\underline
x}}^2}(1-\tilde{V}({\underline x}))\eqno(16)$$

\noindent with

$$\tilde{V}({\underline x}) = \int d^2\ell e^{-i{\underline \ell}\cdot
{\underline x}} V({\underline \ell}).\eqno(17)$$

\noindent Now defining

$$\tilde{N}({\underline x}) = \int d^2\ell \ e^{-i{\underline
\ell}\cdot {\underline x}}{dN({\underline \ell})\over
d^2\ell}\eqno(18)$$

\noindent one easily finds from (6) that

$$N({\underline x}) = \int d^2b dz_0\rho({\underline
b},z_0)\tilde{N}_0({\underline x}) \tilde{f}(z,{\underline
x})\vert_{z={\sqrt{R^2-b^2}}}\eqno(19)$$

\noindent The solution to (14) and (15) is clearly

$$\tilde{f}(z,{\underline x}) = exp\{-{(z-z_0)\over 4\lambda}
{\underline x}^2\tilde{v}({\underline x})\}\eqno(20)$$

\noindent leading to

$$\tilde{N}({\underline x}) = \int d^2b {N_c^2-1\over \pi^2\alpha
N_c{\underline x}^2} (1-exp[-{{\sqrt{R^2-b^2}}\over 2\lambda}
{\underline x}^2\tilde{v}({\underline x})])\eqno(21a)$$

\noindent or

$$\tilde{N}({\underline x})=\int d^2b {N_c^2-1\over \pi^2\alpha
N_c{\underline x}^2}(1-exp[-{2\pi^2{\sqrt{R^2-b^2}}\ {\underline
x}^2\alpha N_c\over N_c^2-1}\rho {x}G({x},{\underline
x}^2)])\eqno(21b)$$

\noindent where we have used (See Eq.4.7 and Appendix B of Ref.12.)

$${\tilde{v}({\underline x})\over \lambda} = {4\pi^2\alpha N_c\over
N_c^2-1} \rho xG(x,{\underline x}^2),\eqno(22)$$

\noindent valid in a logarithmic approximation for small ${\underline
x}^2.$ (Our notation is such that $xG(x,{\underline x}^2)$ is not the
Fourier transform of $xG,$ but rather stands for $xG(x,1/{\underline
x}^2).)$

Eq.(21) is the same result found in Ref.5.  In the limit of very small
${\underline x}^2$ one can keep only the first two terms in the
expansion of the exponential in (21) to arrive at

$$N({\underline x}) = A xG(x,{\underline x}^2)\eqno(23)$$

\noindent which, with the identification of the nuclear gluon
distribution with $N(x_\perp^2),$ gives

$$xG_A(x,{\underline x}^2) = A xG(x,{\underline x}^2).\eqno(24)$$

\noindent At first sight it would seem that (24) no longer holds for
extremely large nuclei at fixed ${\underline x}^2$ where

$$N({\underline x}) \approx {N_c^2-1\over \pi\alpha N_c}\ {R^2\over
{\underline x}^2}.\eqno(25)$$

\noindent However (24) is still correct.  What fails is the
identification of $N({\underline x})$ with $xG_A(x,{\underline x}^2)$
when \ A\ is very large.  The physics here is quite clear.  The
current \ j\ makes a (local) hard interaction with one of the nucleons
in the nucleus and produces a gluon.  The coherence length for this
production is small and so the production cross section, and hence the
nuclear gluon distribution $xG_A(x,Q^2),$ is incoherent among the
nucleons in the nucleus\cite{LLF}.  The quantity which is calculated
in (6) and (21) is the momentum distribution of produced gluons, and
the ${\underline x}$ in (21) is conjugate to the ${\underline \ell}$
of the produced gluon but has no natural connection with the $Q^2$ of
the current when\ A\ is very large.  When\ A\ is not too large, and
the final state interactions do not change the momentum of the
produced gluon too much, {\em and if} one takes a cutoff on
${\underline \ell}^2$ equal to the $Q^2$ of the current, then
$N({\underline x}^2=4/Q^2)$ can be identified with $xG_A(x,{\underline
x}^2)$ and (23) holds. However, when the final state interactions give
a broadening of the transverse momentum distribution
$\Delta{\underline \ell}^2$ which is large compared to $Q^2$ then
$N({\underline x}^2 = 4/Q^2)$ has no relation to the nuclear gluon
distribution.

\subsection{Light-cone gauge calculation of gluon production 
and of the nuclear gluon distribution}

In a light-cone gauge calculation of gluon production one must also
arrive at (21) and (24).  Indeed (21) has already been obtained in
Ref.5 using a light-cone gauge although the interpretation of what
exactly was calculated is, perhaps, not so straightforward.  In this
section, we calculate gluon production in $\eta\cdot A = A_+ =0$
light-cone gauge.  We restrict our discussion to $Q^2 \gg {\underline
\ell}^2.$ We shall find (21) although we shall see that there are no
final state interactions when ${\underline \ell}^2 \ll Q^2.$ The
transverse momentum distribution of the produced gluon now comes from
the wavefunction of the nucleus.

We begin by showing that final state interactions are absent.  The
only unusual aspect of our calculation is the way we choose the $i
\epsilon's$ in the light-cone gauge propagator.  With the momentum as
indicated in Fig.3 our light-cone gauge propagator is

$$D_{\alpha\beta}(k) = {-i\over k^2 +
i\epsilon}[g_{\alpha\beta}-{\eta_\alpha k_\beta\over \eta\cdot
k-i\epsilon} - {k_\alpha\eta_\beta\over \eta\cdot
k+i\epsilon}].\eqno(26)$$

\noindent This choice was used in Ref.11 and implicitly in Refs.4 and
5.  If $\eta\cdot k=k_+$ then the choice of $i\epsilon 's$ in (26)
dictates that a charge at an $x_ -={x_0-x_3\over\sqrt{2}}$ coordinate
$x_-^{(0)}$ gauge rotates all charges having $x_-< x_-^{(0)}$ but does
not gauge rotate charges with $x_-> x_-^{(0)}.$

\subsubsection{Final state interactions are absent}

In order to understand why final state interactions do not appear in
deep inelastic scattering in light-cone gauge it is useful to begin
with the example of gluon production off a quark in one nucleon and
the subsequent rescattering of the gluon off a quark in another
nucleon as illustrated in Fig.4.  The quarks are assumed to be
right-movers and the produced gluon, $\ell,$ is assumed to obey
$\ell_+ \ll p_{i+}.$ The quarks are on-shell before and after the
scattering while the quark having $p_2$ is taken to have its
$x_-$-coordinate to be greater than the $x_-$-coordinate of $p_1$ so
that it is natural to expect a production of the gluon off quark 1 and
a later rescattering off quark 2.  The $+$ symbols along the quark
lines in Fig.4 indicate that the $\gamma_+$ matrices will be dominant,
as usual when considering soft gluon production.

The relevant factors for the graph shown in Fig.4a are

$$G_a={-i(k+q-\ell)_\alpha^\perp\over (\b{k}+\b{q}-{\underline
\ell})^2[(k+q-\ell)_++i\epsilon]}[(k-\ell +
q)_\beta(k-\ell)_\alpha-g_{\alpha\beta}(k-\ell)\cdot(k+q-\ell)]$$

$${i\over 2\ell_-[(k-\ell)_++{(\b{k}-{\underline \ell})^2\over
2\ell_-} - i\epsilon]}\
[g_{\beta\gamma}-{\eta_\beta(k-\ell)_\gamma\over (k-\ell)_++i\epsilon}
-{\eta_\gamma(k-\ell)_\beta\over (k-\ell)_+-i\epsilon}]$$

$$\Gamma_{\gamma\sigma\nu}{-i k_\sigma^\perp\over
\b{k}^2(k_+-i\epsilon)} \epsilon_\nu(\ell).\eqno(27)$$

\noindent We now carry out an integration over $k_+,$ recalling that
$\ell_+ \ll - q_+ \ll p_+. (\ell_+ \ll - q_+$ follows from
$\ell_-=q_-$ and $Q^2 = - 2q_+q_- \gg {\underline \ell}^2$ while $-q_+
\ll p_+$ follows from the fact that we are considering small-x deep
inelastic scattering with $x = -q_+/p_+.$ We take transverse
components of $q_\mu$ to be zero.)  Before doing the $k_+$ integration
we use the Slavnov-Taylor-Ward (STW) identities to drop the
$\eta_\beta(k-\ell)_\gamma$ part of the gluon propagator
$D_{\beta\gamma}$ as this term cancels between the three graphs,
(a-c), of Fig.4.  After this term is dropped only the ${1\over
(k+q-\ell)_++ i\epsilon}$ has a singularity in the lower half $k_+-$
plane.  We can do the $dk_+$ integration by distorting the contour in
the lower half $k_+-$ plane so long as the rest of the integrand, the
other terms in $G_a,$ vanish for large values of $k_+.$ And if that is
the case the net result will be small because the final evaluation is
done at $k_+=(\ell - q)_+ \approx - q_+$ so that inverse powers of
$k_+$ correspond to higher inverse powers of $Q^2$ as compared to the
case of no rescattering. In order to count the $k_+-$ factors in $G_a$
it is convenient to write

$$g_{\beta\gamma}- {\eta_\gamma(k-\ell)_\beta\over
(k-\ell)_+-i\epsilon} = \eta_\beta\bar{\eta}_\gamma +
g_{\beta\gamma}^\perp + {\eta_\gamma
(\eta_\beta\ell_--(k-\ell)_\beta^\perp)\over (k-\ell)_+ -
i\epsilon}.\eqno(28)$$

\noindent Then in order to compensate for the $k_+-i\epsilon$ and
$(k-\ell)_+ +{(\b{k}-{\underline \ell})^2\over 2\ell_-}-i\epsilon$
denominators, not including the possible denominator in (28), one
would need a $k_+^2$ coming from the numerator terms in order not to
have a small result.  This means that one must get a single factor of
$k_+$ (or $q_+$) each from the $\Gamma_{\gamma\sigma\nu}-$ term and
from the $[ ]_{\beta\alpha}-$ term in (27).  In order to get a $k_+-$
factor from $\Gamma_{\gamma\sigma\nu}$ an accompanying
$\bar{\eta}_\gamma$ , $\bar{\eta}_\sigma$ or $\bar{\eta}_{\nu}$ must
appear.  But $\bar{\eta}_\sigma$ clearly gives zero from (27) while
$\bar{\eta_\gamma}$ gives zero, or an extra $k_+$ in the denominator,
from (28) so we are left with $\bar{\eta}_\nu$ as the only
possibility.  Here we can simplify our discussion by choosing
$\bar{\eta}\cdot \epsilon = 0.$ This is perhaps an unusual choice to
go with $\eta\cdot A = 0$ light-cone gauge but, after all, one is free
to choose the $external$ polarizations in any convenient basis.  With
the choice ${\bar{\eta}}\cdot \epsilon = 0$ we find

$$\int dk_+ G_a = 0\eqno(29)$$

\noindent (We could arrive at this result somewhat more directly.
After observing that there is sufficient convergence to do the $k_+-$
integration by distorting in the lower half plane we could use current
conservation in the $\alpha$-index of the $q-$vertex along with
$(k+q-\ell)_\alpha^\perp\approx (k+q-\ell)_\alpha$ to get (29).
However, it does not seem possible to generalize this procedure to all
higher order graphs.)

The choice $\bar{\eta}\cdot\epsilon = 0$ also makes graphs\ b\ and\ c\
of Fig.4 equal to zero because one cannot have a $\gamma_+$ matrix
with this polarization vector.  Thus we are left only with the graph
shown in Fig.4d and we must also evaluate it with
$\bar{\eta}\cdot\epsilon = 0.$ Now

$$G_d = {-i(k-\ell + q)_\alpha^\perp\over (\b{k}-{\underline \ell} +
\b{q})^2[(k + q-\ell)_++ i\epsilon]}\ {-i k_\sigma^\perp\over
\b{k}^2(k_+-i\epsilon)}
\epsilon_\nu\Gamma_{\alpha\sigma\nu}^q\eqno(30)$$

\noindent where

$$\Gamma_{\alpha\sigma\nu}^q =
g_{\alpha\sigma}(\ell-q-2k)_\nu-q_{\alpha\nu}(2\ell-q-k)_\sigma +
g_{\nu\sigma}(k+\ell)_\alpha.\eqno(31)$$

\noindent Again, in doing the $k_+$ integration we will end up with a
$q_+$ in the denominator, giving a small result, unless one can take a
$k_+,$ or a $q_+,$ from $\Gamma$ as given in (31).  But, taking a
$k_+,$ or a $q_+,$ from $\Gamma$ will result in a factor of
$\bar{\eta}_\nu $, $\bar{\eta}_\sigma$ or $\bar{\eta}_\alpha$ each of
which give zero in (30).

We have seen that the graphs of Fig.4 give a small contribution. (We
have reached the conclusion for ${\underline \ell}^2 \ll Q^2.$ Of
course we would have found a small result for ${\underline {\ell}}^2$
large independently of the size of $Q^2,$ but that is no different
than in covariant gauge.  The fact that final state interactions are
suppressed for large $Q^2,$ and with ${\underline \ell}^2 \ll Q^2,$ is
different than what we found in covariant gauge.)  In a moment we
shall give a physical argument explaining why this is not unexpected.
However, before describing our physical argument consider the high
energy gluon quark scattering amplitude at order $\alpha$ in $A_+=0$
light-cone gauge and in a frame where the quark is right-moving and
the gluon left-moving. The graphs are shown in Fig.5.  If we choose
$\bar{\eta}\cdot \epsilon = 0$ for the initial and final gluons, then
only the first graph in Fig.5 contributes. The essential factors in
this graph are

$$G = \tilde{u}(p-k)\gamma_+ u(p) {i\over \b{k}^2}\
{k_\alpha^\perp\over k_+} \Gamma_{\alpha\beta\gamma}
\epsilon_\gamma^{(\lambda_i)}\bar{\epsilon}_\beta^{(\lambda_f)}.\eqno(32)$$

\noindent The on-shell conditions $(\ell-k)^2=0$ and $(p-k)^2=0$ give
$k_+={{\underline \ell}^2-({\underline \ell}-\b{k})^2\over 2\ell_-},
k_- = - \b{k}^2/2p_+.$ In order to see how the high energy limit comes
about it is useful to write

$$k_\alpha^\perp = k_\alpha -\bar{\eta}_\alpha k_+-\eta_\alpha k_-.\eqno(33)$$

\noindent $k_\alpha$ gives zero while acting in (32) while
$\eta_\alpha$ gives a small factor.  Using

$$\bar{\eta}_\alpha \Gamma_{\alpha\beta\gamma}
\epsilon_\gamma^{\lambda_i)}\bar{\epsilon}_\beta^{(\lambda_f)} =
2\ell_-\epsilon^{(\lambda_i)}\cdot
\bar{\epsilon}^{(\lambda_f)}\eqno(34)$$

\noindent gives

$$G = {2 i s\over \underline{k}^2} \ {\underline \epsilon}^{(\lambda_i)}\cdot
{\underline{\bar{\epsilon}}}^{(\lambda_f)}.\eqno(35)$$

\noindent Of course, the result emerges directly from an evaluation of
(32) without explicitly invoking current conservation, (33).  In a
frame where $\ell_-$ is large the high energy growth, the factor of
$s$ in (35), comes partly from the smallness of $k_+$ in the
denominator in (32).  In order to get a small denominator it is
crucial to have a long region of integration over $x_-,$ the variable
conjugate to $k_+.$ Of course the quark and gluon are far apart when
$x_-$ is large, but in axial gauge important interactions can happen
in a non-causal manner.  In our choice of light-cone gauge, (26), the
factor ${1\over k_+}$ in (32) comes from times long before the quark
and gluon have reached each other.

We are now in a position to see why final state interactions cannot be
important in single gluon production in light-cone gauge in deep
inelastic scattering.  In Fig.4a and its higher order counterparts a
final state interaction occurs much like the one we have just
considered.  In order for this final state interaction to be important
one would need a trapping of the $k_+-$ integration contour at values
of $k_+$ on the order of ${\underline \ell}^2/2\ell_-$ but from (27)
we see that we can distort the $k_+$-contour so that $k_+$ is always
of size $q_+ = - {Q^2\over 2\ell_-}.$ Because $Q^2$ is a large
parameter $k_+$ is not forced to be small, and thus final state
interactions are weak.  We can say this slightly differently.  The
highly off-shell current has a lifetime $\Delta x_-\approx {2q_-\over
Q^2}$ and this is the limiting $x_-$ over which integrations can be
done.  This $\Delta x_-$ is not large enough to generate the small
light-cone denominators necessary to have final state interactions, at
least so long as the target is not too long compared to its
longitudinal momentum, that is so long as ${\underline
\ell}^2{<<}Q^2.$

\subsubsection{Gluon production in deep inelastic scattering}

We begin by considering deep inelastic scattering in a nucleus where
two separate nucleons are involved in the scattering.  The two classes
of graphs are illustrated in Fig.6.  The graphs of Fig.6a correspond
to inelastic scattering off both nucleon 1 and off nucleon 2 while
graphs in the class of Fig.6b correspond to an inelastic reaction with
nucleon 1 and an elastic reaction with nucleon 2.  We emphasize that
in our frame where the nucleons are right-movers and the current\ q\
has $q_- \gg q_+$ the $x_-$- coordinate of nucleon 1 is less than the
$x_- -$ coordinate of nucleon 2.  The interactions shown in Fig.6 are
nonzero only when the propagator of the $k-$line is taken to be ${i
\eta_\beta k_\alpha\over [k^2+ i\epsilon][k_+-i\epsilon]}$ and
similarly for the $(k-\Delta)-$line.  (We suppose
$\Delta_\perp=\Delta_-=0$ while $\Delta_+$ is integrated freely over
values large compared to $\ell_+$ as is appropriate for a low x
collision where the produced gluon has a coherence length large
compared to the length of the nucleus.)  Our gauge choice, as
indicated in (26), allows the gluon field of a nucleon to have a pure
gauge part which extends in the negative $x_ -$direction starting from
that nucleon. As shown in detail in Ref.11 this field gauge rotates
the field of nucleons having smaller values of $x_-.$ Thus, the field
of nucleon 2 rotates the field coming from nucleon 1.

Now let us evaluate explicitly the graphs in Fig.6 starting with those
in Fig.6a.  The $k-$line of that figure goes from nucleon 2 to nucleon
1 or to the field coming from nucleon 1.  Because of the $k_\alpha$ at
the end of the $k-$line we can use the STW identities to evaluate
these contributions. This has already been done in Ref.11 and here we
need only outline the procedure and put in the normalizations
appropriate for our problem.  In doing the $k_+-$ integration we
distort over the ${1\over k_+-i\epsilon}$ pole if the light-cone
propagator which sets $k_+=0.$ Since $k_-=0$ compared to $\ell_-$ the
$k-$line, which by the STW identities eliminates the
$(\ell-q)$-propagator, effectively inserts a transverse momentum $k$
just before the attachment of the current q.  In terms of formulas

$$\int dk_+ {k_\alpha\over k_+-i\epsilon} \Gamma_{\sigma\alpha\rho}
D_{\rho\mu}(\ell-q)v_{\mu\lambda}= - 2\pi g v_{\sigma\lambda} + \cdot
\cdot \cdot \eqno(36)$$

\noindent where $\Gamma_{\sigma\alpha\rho}$ is the triple gluon
vertex, $D_{\rho\mu}$ the gluon propagator and $v_{\mu\nu}$ the vertex
of the current\ $q$\ with the gluon lines $\ell-q$ and $\ell.$ The
omitted terms in (36) cancel with gluon, $k_\alpha,$ attachments into
the lower {\em blob} of nucleon 1.  In exactly the same way, and
integrating over $\Delta_+,$ the gluon line $k-\Delta$ also eliminates
the $\ell-q$ line in the complex conjugate amplitude.

Thus, the STW identities allow one to calculate the effect of nucleon
2 on the field coming from nucleon 1.  We may write the gluon
production amplitude in terms of the scattering of the current off the
gluon $(\ell-q),$ as illustrated in Fig.7, times the unintegrated
gluon distribution coming from nucleon 1 as modified by nucleon 2.
Rather than immediately writing the answer for the two nucleon case we
instead proceed to the general case using the ``classical'' field
calculated in Ref.4 to generate the tree graphs contributing to the
unintegrated gluon distribution coming from an arbitrary number of
nucleons in the nucleus.  We change notation slightly from Ref.4 to
better match the problem at hand.  Write $A_\mu^\perp({\underline x},
x_-) = \sum_{a=1}^{N_c^2-1} T^aA_\mu^{a\perp}({\underline x}, x_-)$ in
light-cone gauge as

$$A_\mu^\perp({\underline x}, x_-) = \int S({\underline x},b_-) T^a
S^{-1} ({\underline x},b_-) \nabla_\mu^\perp \ln [\vert{\underline
x}-{\underline b}\vert\mu]{\hat{\rho}}^a({\underline b},b_-)\theta(b_-
- x_-) d^2bdb_- \eqno(37)$$

\noindent with

$$S({\underline x}, x_-) = P \exp \left\{ {{i g T^a}} \int \ln
[\vert{\underline x}-{\underline b}\vert\mu]{\hat{\rho}}^a({\underline
b},b_-)\theta(b_-- x_-)d^2 bdb_- \right\} \eqno(38)$$

\noindent where P path orders the $b_--$integration with terms having
smaller values of $b_-$ coming more to the right.  ${\hat{\rho}}^a$ is
a color charge operator normalized according to

$$<{\hat{\rho}}^a({\underline b},
b_-){\hat{\rho}}^{a^\prime}({\underline b},^\prime b_- ^\prime)> =
{\rho({\underline b},b_-)\over N_c^2-1} \delta({\underline
b}-{\underline b}^\prime)\delta(b_- -b_-^\prime)\delta_{aa^\prime}
Q^2{\partial\over \partial Q^2} xG(x,Q^2)\eqno(39)$$

\noindent with $\rho({\underline b},b_-)$ the normal nuclear density,
in our boosted frame, obeying

$$\int d^2b db_-\rho({\underline b},b_-) = A.\eqno(40)$$

\noindent $\mu$ in (37) and (38) is an infrared cutoff which will
disappear from physical quantities.  We note that $Q^2{\partial\over
\partial Q^2} xG(x,Q^2)$ is independent of $x$ and $Q^2$ in our
approximation.  The expectation indicated in (39) is an expectation in
the nuclear wavefunction and we suppose that the pairwise correlation
indicated there is the only nontrivial correlation.  This is a natural
assumption for a nucleus having weak nucleon-nucleon correlations.
Define a Fourier transformed field by

$$\b{A}({\underline k},k_+) = \int {d^2x\over 2\pi} \int dx_-
e^{i(k_+-i\epsilon)x_-- i\underline{k}\cdot {\underline x}}
{\underline{A}}({\underline x}, x_-).\eqno(41)$$

\noindent Then the main formula which we shall need in order to
evaluate ${dN\over d^2\ell}$ is

$$\pi{dN\over d^2\ell} = - 2Tr<A_\mu^\perp({\underline
\ell},\ell_+-q_+) A_\nu^\perp(-{\underline \ell},-\ell_+
+q_+)>[g_{\mu\nu}^\perp{(Q^2-{\underline \ell})^2\over
4}-\ell_\mu^\perp\ell_\nu^\perp Q^2]{1\over \ell_-^2}\eqno(42)$$
\bigskip

\noindent where the term in the brackets on the right-hand side of
(42) is given by the graph in Fig.7.  The $A_\mu$ in (42) is the same
as given in (37) and represents the field of the nucleus which the
source \ j\ interacts with.  Eq.(42) does not have any final state
interactions, interactions of the produced gluon with the nucleus
after the action of \ j,\ which we have previously argued are small.
As $Q^2 \gg {\underline \ell}^2$ and $\ell_+ \ll -q_+$ then taking the
leading $Q^2$ term in Eq.(42) and using Eq.(18) yields

$$\tilde{N}({\underline x}^2) = -{2\over \pi} \int
d^2bTr<A_\mu^\perp({\underline b})A_\mu^\perp({\underline b} +
{\underline x})>.\eqno(43)$$

\noindent Plugging in $A_\mu^\perp$ from Eq.(37) one finds

$$\tilde{N}({\underline x}^2) = - {2\over \pi} \int d^2b \int
d^2b^\prime db_-^\prime d^2b^{\prime\prime} db_-^{\prime\prime}$$

$$ <{{\underline b}-{\underline b}^\prime\over \vert{\underline
b}-{\underline b}^\prime\vert^2}\cdot {{\underline b}+{\underline x} -
{\underline b}^{\prime\prime}\over\vert{\underline b}+{\underline
x}-{\underline b}^{\prime\prime}\vert^2} {\hat{\rho^a}}({\underline
b}^\prime,b_-^\prime){\hat{\rho}}^b ({\underline
b}^{\prime\prime},b_-^{\prime\prime})$$

$$Tr[S({\underline b},b_-^\prime)T^aS^{-1}({\underline
b},b_-^\prime)S({\underline b}+ {\underline
x},b_-^{\prime\prime})T^aS^{-1}({\underline b}+ {\underline x},b_-
^{\prime\prime})]>.\eqno(44)$$

\noindent Here we can assume that the path ordered $b_-$-integration
in the $S({\underline x},x_-)$-matrices [see Eq.(38)] is performed up
to some point close to $x_-,$ but excludes this point and its
immediate vicinity.  This statement is equivalent to what was referred
to as dropping the {\em last} nucleon in Ref.4.  Using these
assumptions we can independently calculate the correlation of the two
color densities in (44).  Employing (39) and integrating over
$d^2b^\prime$ we arrive at

$$\tilde{N}({\underline x}^2) = 4{\rho_{rel}\over N_c^2-1} \ell
n(\vert{\underline x}\vert \mu) Q^2{\partial\over \partial Q^2}
xG(x,Q^2)$$

$$ \int d^2bdb_-^\prime<Tr[S({\underline
b},b_-^\prime)T^aS^{-1}({\underline b}, b_-^\prime)S({\underline
b}+{\underline x},b_-^\prime)T^aS^{-1}({\underline b}+ {\underline
x},b_-^\prime)]>,\eqno(45)$$

\noindent where $\mu$ is some infrared cutoff, and we have assumed
that the normal nuclear density is uniform throughout the nucleus and
in the boosted frame is given by $\rho_{rel}.$ Rewriting (38)
according to the definition of a path-ordered product we obtain

$$S({\underline b},b_-) = {\prod_ i}[1 + igT^a
{\hat{\rho}}^a(\b{y},y_-) \ell n(\vert{\underline
b}-\b{y}\vert\mu)d^2y\Delta y_{i-}$$

$$-(1/2)g^2 T^a T^b {\hat{\rho}}^a(\underline{y},y_-){\hat{\rho}}^b
(\underline{y}^\prime,y_-^\prime) \ell n(\vert{\underline
b}-\underline{y}\vert\mu)\ell n(\vert{\underline
b}-\underline{y}^\prime\vert\mu)$$

$$ d^2y \Delta y_{i-}d^2y^\prime\Delta y_{i-}^\prime],\eqno(46)$$

\noindent where the product goes along the $y_--$axis, such that
smaller $y_-$ corresponds to greater label i.  For a combination of
matrices like $S^{-1}({\underline b},b_-^\prime) S({\underline b} +
{\underline x},b_-^\prime)$ we start by considering the last
(rightmost) term of the product given by (46) for $S^{-1}({\underline
b},b_-^\prime)$ and the first (leftmost) term in a similar product for
$S({\underline b}+{\underline x},b_-^\prime).$ Both terms include the
same interval of the $y_-$ integration, the closest to $b_-^\prime.$
Since all $S-$matrices in (45) are taken at the same longitudinal
coordinate $b_-^\prime$, then taking these terms for both pairs of $S-
$matrices in (45) we observe that nothing else in (45) depends on the
longitudinal coordinates in this interval.  Therefore we can average
these terms independently of the rest of the expression in (45), as
well as do the transverse integrations. Taking only the two-density
correlation terms (We throw away the higher order correlations since
they have more powers of \ $g,$\ which goes beyond the classical
approximation\cite{YK}) and making use of (39) we end up with

$$<Tr[S({\underline b},b_-^\prime)T^aS^{-1}({\underline
b},b_-^\prime)S({\underline b}+{\underline
x},b_-^\prime)T^aS^{-1}({\underline b}+ {\underline x},b_-^\prime)]>$$
 $$=\left(1-g^2{\pi\rho_{rel}N_c{\underline x}^2\over 4(N_c^2-1)}
xG(x, 1/{\underline x}^2) \Delta y_-\right)$$

$$ <Tr[S({\underline b},b_-^\prime - \Delta y_-)T^aS^{-1}({\underline
b},b_-^\prime - \Delta y_-) S({\underline b}+{\underline x},b_-
^\prime - \Delta y_-)T^aS^{-1}({\underline b}+{\underline x},
b^\prime_- - \Delta y_-)]>,\eqno(47)$$

\noindent where $\Delta y_-$ is the absolute value of the change in
longitudinal coordinate.  The transverse integration $d^2y$ was done
assuming that the nucleus is infinite in the transverse direction,
which is a reasonable approximation for a large nucleus.  In arriving
at (47) we have neglected QCD evolution in xG.  Continuing this
procedure of picking small intervals along the $y_--$axis and taking
the limit of $\Delta y_-\to 0$ shows that the trace in (47) is equal
to an exponential function

$$<Tr[S({\underline b},b_-^\prime)T^aS^{-1}({\underline
b},b^\prime_-)S({\underline b}+{\underline x},
b^\prime_-)T^aS^{-1}({\underline b}+{\underline x},b_-^\prime)]> =$$
$$C_FN_c exp\left(-g^2{\pi\rho_{rel}N_c{\underline x}^2\over
4(N_c^2-1)} xG(x,1/{\underline x}^2)(b_-^\prime +
b_{0-}^\prime)\right),\eqno(48)$$

\noindent with $\pm b_{0-}^\prime$ the\ upper\ (lower)\ limit\ of\ the
$y_-$ integration in the $S-$matrices in (47).  Plugging this back in
(45), performing the $db_-^\prime$ integration and defining the
nuclear density in the center of mass frame $\rho =
\rho_{rel}/\gamma{\sqrt{2}},$ with \ $\gamma\ $ the Lorentz
contraction factor provides us with

$$\tilde N({\underline x}^2) = {N_c^2-1\over \pi^2\alpha
N_c{\underline x}^2} \int d^2b\left[1- exp\left(-{2\pi^2\alpha
N_c{\underline x}^2{\sqrt{R^2-b^2}}\over N_c^2-1}\rho x G(x,
1/{\underline x}^2)\right)\right],\eqno(49)$$

\noindent which is exactly formula (21b).  Thus one indeed is able to
calculate the produced gluon spectrum from the ``classical'' field
given in (37) and (38) and this accounts for the final state
interactions which are present in covariant gauge.

\section{The produced gluon distribution in nucleon-nucleus collisions}

In this section we shall calculate the spectrum of gluons produced in
the scattering of a nucleon off a large nucleus.  If the nucleus is
right-moving and the nucleon left-moving then we suppose the centrally
produced gluon momentum, $\ell,$ is such that $\ell_-$ is much less
than the minus component of the nucleon momentum and that $\ell_+$ is
sufficiently less than the momentum of a nucleon in the nucleus so
that the gluon is coherent over the longitudinal extent of the
nucleus.  We begin our discussion with gluon production in
nucleon-nucleon collisions.

\subsection{Soft gluon production in nucleon-nucleon collisions}

We choose $A_-=0$ gauge and a center of mass system to describe gluon
production.  If we choose the outgoing gluon polarization such that
$\eta\cdot\epsilon = \epsilon_+ =0$ then we need only consider the
graph shown in Fig.8 for the gluon production amplitude.  The terms
occurring at the 3-gluon vertex in that graph are

$$V^{(\lambda)}={1\over
k_-(\ell-k)_-}(\ell-k)_\gamma^\perp
\bar{\eta}_\alpha\epsilon_{\beta}^{(\lambda)}[-g_{\alpha\gamma}(2k-\ell)_\beta+
g_{\beta\alpha}(k+\ell)_\gamma-g_{\gamma\beta}(2\ell-k)_\alpha]\eqno(50)$$

\noindent where $p_{1+}$ and $p_{2-}$ are the large components of
$p_1$ and $p_2$ respectively.  It is straightforward to check that

$$V^{(\lambda)}={2\over k_-{\underline \ell}^2}[{\underline
\ell}^2{\underline \epsilon}^{(\lambda)}\cdot({\underline
\ell}-\underline{k})-({\underline \ell}^2- \underline{k}^2){\underline
\epsilon}^{(\lambda)}\cdot{\underline \ell}]\eqno(51)$$

\noindent and that

$$\sum_{\lambda=1}^2 V^{(\lambda)} V^{(\lambda) *} = {4
\underline{k}^2 ({\underline \ell}-\underline{k})^2\over
k_-^2{\underline \ell}^2}.\eqno(52)$$

\noindent There is an $\bar{\eta}_\rho$ at the $\rho-$vertex in Fig.8
while there is a $ k_\sigma^\perp$ at the $\sigma-$vertex.  Writing

$$k_\sigma^\perp = k_\sigma - \eta_\sigma k_--\bar{\eta}_\sigma
k_+\eqno(53)$$

\noindent and using the fact that $k_\sigma$ gives zero by current
conservation while $\bar{\eta_\sigma}$ gives a very small result one
can replace $k_\sigma^\perp$ by $-\eta_\sigma k_-.$ The $k_-$ will
cancel the $k_--$ denominator in (51) (and in (52)) so that in each
case the vertices $\rho$ and $\sigma$ are multiplied by the
appropriate factors for the cross section to be given by the product of
the unintegrated gluon distributions of nucleons $p_1$ and $p_2.$ The
result is\cite{EML}-\cite{XG}.

$${d\sigma_0\over d^2\ell dy} = {4\alpha N_c\over (N_c^2-1){\underline
\ell}^2} \int d^2k {\partial x_1G(x_1,k^2)\over \partial k^2}\
{\partial x_2G(x_2,({\underline \ell}-\underline{k})^2)\over \partial
({\underline \ell}-\underline{k})^2}\eqno(54)$$

\noindent where $x_1x_2 s = {\underline \ell}^2$ and\ y\ is the
rapidity of the produced gluon.  Of course (54) is difficult to take
literally.  If ${\underline \ell}^2$ is not large there is no hard
scale and the unintegrated gluon distributions do not have much
meaning.  On the other hand if ${\underline \ell}^2$ is large then the
cross section is better expressed as the product of two gluon
distributions times a hard scattering cross section.  Nevertheless,
(54) is useful as a normalization and for comparison with the
equations we are now going to derive for gluon production in
nucleon-nucleus collisions.  It also follows in a BFKL
approximation\cite{VDD,KJE}.

\subsection{Central region gluon production in nucleon-nucleus collisions}

It is not too difficult to calculate the cross section for producing a
gluon of transverse momentum ${\underline \ell}$ in the collision of a
nucleon with a large nucleus.  It is convenient to view the
calculation in the rest system of the nucleus.  We suppose the
longitudinal momentum of the gluon is large enough so that the gluon
is coherent over the length of the nucleus\cite{LLF}.  We begin by
considering gluon radiation when a quark scatters on a nucleus.  It
will then be a simple matter to extend the result to an incoming
nucleon.  It is convenient to work in $\bar{\eta}\cdot A = A_-=0$
gauge and with $\bar{\eta}\cdot\epsilon = 0$ where we choose the
incoming quark to have a large minus component of momentum.  Suppose
the quark reaches the nucleon at light-cone time $\tau = x_-=0.$ Then
we separate the gluon radiation cross section into four terms
depending on the time $\tau_1$ that the gluon is emitted from the
quark in the amplitude and on the time $\tau_2$ that the gluon is
emitted from the quark in the complex conjugate amplitude.  The four
cases are:

 $$(a)\ \ \ \ \ \tau_1 > 0, \tau_2 > 0\eqno(55a)$$
 $$(b)\ \ \ \ \ \tau_1 < 0, \tau_2 > 0\eqno(55b)$$
 $$(c)\ \ \ \ \ \tau_1 > 0, \tau_2 < 0\eqno(55c)$$
 $$(d)\ \ \ \ \ \tau_1 < 0, \tau_2 < 0\eqno(55d)$$
 
 \noindent Since the coherence time of the gluon is assumed to be much
bigger than the radius of the nucleus we can assume that the quark, or
the quark-gluon system, passes over the nucleus instantaneously
compared to the magnitudes of the times of emission $\tau_1$ and
$\tau_2.$ Also once we separate contributions into definite times we
are in effect dealing with light-cone perturbation theory rather than
with Feynman graphs though in many graphs the orderings (55) simply
tell whether the gluon is emitted before or after the quark, or
quark-gluon system, interacts with the nucleus.  The cross sections
coming from the regions (55) are given by the expressions

$${d\sigma^{(a)}\over d^2\ell dy} = {1\over \pi} \int d^2b d^2x_1 d^2
x_2 {1\over 4\pi^2} {\alpha C_F\over \pi} {{\underline x}_1\cdot
{\underline x}_2\over x_1^2x_2^2} e^{i{\underline \ell}({\underline
x}_1-{\underline x}_2)}\eqno(56)$$

$${d\sigma^{(b+c)}\over d^2\ell dy} = {-1\over \pi} \int d^2b d^2x_1
d^2x_2 {1\over 4\pi^2} {\alpha C_F\over \pi}\ {{\underline x}_1\cdot
{\underline x}_2\over x_1^2x_2^2}\cdot$$
$$\left( \exp[-{{{\underline x}_1^2 \tilde{v} {\sqrt{R^2-{\underline
b}^2}}}\over{2\lambda}}] + \exp[-{{{\underline
x}_2^2\tilde{v}{\sqrt{R^2-{\underline b}^2}}}\over{2\lambda}}] \right)
e^{i{\underline \ell}({\underline x}_1-{\underline x}_2)}\eqno(57)$$

$${d\sigma^{(d)}\over d^2\ell dy} = {1\over \pi}\int d^2b d^2x_1d^2x_2
{1\over 4\pi^2} {\alpha C_F\over \pi}\ {{\underline x}_1\cdot
{\underline x}_2\over x_1^2x_2^2} exp[{-({\underline x}_1-{\underline
x}_2)^2\tilde{v}{\sqrt{R^2-{\underline b}^2}}\over 2\lambda}]
e^{i{\underline \ell}\cdot({\underline x}_1-{\underline
x}_2)}\eqno(58)$$

\noindent In the above we take the transverse coordinate of the quark
to be $\b{o}$ while ${\underline x}_1$ and ${\underline x}_2$ are the
transverse coordinates of the gluon in the amplitude and complex
conjugate amplitude respectively. We use a shorthand notation where
${\underline x}^2\tilde{v}$ means ${\underline
x}^2\tilde{v}({\underline x}^2)$ with $\tilde{v}, \lambda, R$ and
${\underline b}$ as previously used in Sec.2.  Let us now see how (56)
to (58) come about.

In (56) the factor ${1\over 4\pi^2}\ {\alpha C_F\over \pi}
{{\underline x}_1 \cdot {\underline x}_2\over x_1^2x_2^2}$ is just the
product of the coordinate space gluon emission amplitude times the
complex conjugate amplitude.  What may seem surprising in (56) is that
there is no trace of the nucleus! But this is easy to understand.
What passes through the nucleus is a high energy quark.  The quark
does not change its transverse coordinate nor does it lose a
significant amount of energy as it passes through the nucleus.
(Technically, elastic quark-nucleon scatterings and inelastic
quark-nucleon scatterings, where the nucleon breaks up, cancel.)  What
is unusual in (56) is that we are attributing a contribution to the
cross section to the disconnected graph where the quark freely passes
through the nucleus.  The Feynman diagram contribution to this process
would be zero by energy conservation.  However, (56) corresponds to
$\tau_1$ and $\tau_2$ greater than zero so we do not yet have energy
conservation.  There are other terms in (57) and (58) which can be
viewed as cancelling this ``gluon radiation without interaction''
term.  Indeed these are the $v$-independent terms of (57) and (58).
However, the grouping of terms as given in (56)-(58) is convenient for
our purposes.

Now turn to (58).  The exponential term in (61) reflects multiple
scattering of the gluon as it passes through a length $\Delta z =
2{\sqrt{R^2-{\underline b}^2}}$ of nuclear matter.  This factor is
identical to that given by (20) with $z-z_0$ taken to be
$2{\sqrt{R^2-{\underline b}^2}}.$ In this case it is the quark-gluon
system which passes through the nucleus both in the amplitude and in
the complex conjugate amplitude.  Interactions of the quark with the
nucleons in the nucleus cancel between real and virtual (production
and elastic scattering) terms.

The expression (57) corresponds to a quark-gluon system passing
through the nucleus in the amplitude and a quark passing through the
nucleus in the complex conjugate amplitude along with a term where the
amplitude and complex conjugate amplitude terms are exchanged.  In
Appendix A we outline how the exponential factors, involving
${\underline x}_1^2$ and ${\underline x}_2^2,$ come about.

It is straightforward to evaluate (56)-(58) in the approximation of
neglecting the ${\underline x}$-dependence in $\tilde{v}.$ One finds

$${d\sigma^{(a)}\over d^2\ell dy}= \int {\alpha C_F\over \pi^2}
{1\over \ell^2} d^2b\eqno(59)$$

 $${d\sigma^{(b+c)}\over d^2\ell dy} = - 2 \int {\alpha C_F\over
\pi^2} {[1-e^{-\ell^2\over<\ell^2>}]\over \ell^2} d^2b\eqno(60)$$

$${d\sigma^{(d)}\over d^2\ell dy} = \int d^2b {\alpha C_F\over
\pi^2}{e^{-\ell^2\over <\ell^2>}\over <\ell^2>}$$
$$\{\ell n[{<\ell^2>L^2\over 4}]-\Gamma(0,-{\ell^2\over <\ell^2>}) -
\ell n {-\ell^2\over <\ell^2>}\}\eqno(61)$$

\noindent where\cite{RBY}

$$<\ell^2> = <{\underline \ell}^2({\underline b})> =
2{\sqrt{R^2-{\underline b}^2}}\ {\tilde{v}(<\ell^2>)\over
\lambda}.\eqno(62)$$

\noindent L is an infrared cutoff and $\Gamma(n,z)$ is the incomplete
$\Gamma$- function.  In arriving at (61) we have used

$$\int_0^{2\pi} d\phi(\b{y}) {{\underline x} + \b{y}\over ({\underline
x} + \b{y})^2} = 2\pi \Theta(x-y){{\underline x}\over x^2}\eqno(63)$$

\noindent and

$$\int_0^{2\pi} d\phi(\b{y}) {1\over \vert{\underline x} +
\b{y}\vert^2} = {2\pi\over \vert x^2-y^2\vert}\eqno(64)$$

\noindent with ${\underline x}$ and $\underline{y}$ in (63) and (64)
related to $\underline{x}_1$ and ${\underline x}_2$ in (58) by
${\underline x} = \underline{x}_1-{\underline x}_2,
\underline{y}={\underline x}_2$. The second plus third terms in the
brackets on the right-hand side of (61) are non-logarithmic when
$\ell^2 \ll <{\underline \ell}>^2.$ When $\ell^2 \gg <\ell^2>$ this
term exactly cancels similar terms in (59) and (60).  Thus in the
logarithmic approximation we keep only the first term on the
right-hand side of (61) and we understand that (59) holds only for
$\ell^2 \ll <\ell^2>.$

We can now go to the case of an incident nucleon by making the
replacements

$${\alpha C_F\over \pi}\ {1\over \ell^2} \longrightarrow
{\partial\over \partial\ell^2} xG(x,\ell^2).\eqno(65a)$$

$${\alpha C_F\over \pi} \ell n[<\ell^2> L^2] \rightarrow x
G(x,\ell^2).\eqno(65b)$$

\noindent In the logarithmic approximation the (b) + (c) contribution
can be neglected and one finds

$${d\sigma\over d^2\ell dy} = {1\over \pi} \int
d^2b\left[{\partial\over \partial\ell^2}\ xG(x,\ell^2) + xG(x,
<\ell^2>)\ {\exp \left( -{\ell^2 \over <\ell^2>} \right) \over
<\ell^2>}\right]\eqno(66)$$

\noindent with the understanding that the first term on the right-hand
side of (66) can only be used when $\ell^2 \ll <\ell^2>.$ We note that

$${d\sigma\over dy} = \int {d\sigma\over d^2\ell dy} d^2\ell = 2 \int
d^2b\ xG(x, <\ell^2>)\eqno(67)$$

There is a simple physical interpretation of (66) and (67).  When an
incident nucleon passes through a nucleus all the gluons in the
nucleon's wavefunction having $\ell^2{<}\ <{\underline \ell}^2>$ are
freed during the collision and the final state interactions broaden
their distribution.  This is the second term on the right-hand side of
(66) and one half of (67).  The nucleon remnants emerging from the
nucleus have lost their gluon cloud. In rebuilding that cloud there is
further gluon emission, but this gluon emission has no transverse
momentum broadening.  This is the first term on the right-hand side of
(66) and the other half of (67).

\subsection{Gluon production in nucleon-nucleus collisions 
without final state interactions}

In this section we describe exactly the same process as in the
previous section.  However, now we choose $A_+=0$ gauge with
$\epsilon_-=0.$ Our object is to show that, as for our deep inelastic
scattering process of Sec.2, the gluon production cross section can be
described completely in terms of the interaction of the ``classical''
fields of the incoming nucleon and of the incoming nucleus without any
final state interactions.

It is relatively straightforward to derive a general formula for
${d\sigma\over d^2\ell dy}$ in terms of classical fields associated
with the colliding nucleon and nucleus.  To see how this comes about
we discuss in detail the scattering of a left-moving nucleon on two
right-moving nucleons.  We begin with the graphs shown in Fig.9 where
we suppose that nucleons 1 and 2 are right-moving and that the $x_--$
coordinate of nucleon 1 is less than that of nucleon 2, that is,
nucleon 1 is ahead of nucleon 2.  We assume these nucleons are
separated by a longitudinal distance large compared to $1/\wedge$ in
their common rest system.  Then the interaction in Fig.9a must be a
gauge term with the effective part of the k-propagator being
${i\eta_\beta k_\alpha\over \underline{k}^2[k_+-i\epsilon]}.$ The phase factor
$e^{ik_+(x_2-x_1)_-}$ dictates that one distort the contour in the
upper-half plane with the propagator becoming ${-2\pi\over \underline{k}^2}
\delta(k_+)k_\alpha.$ Similarly for the graph in Fig.9b one can again
distort the propagator in the upper half $k_+-$plane getting an
identical result for the $k-$ propagator.  One now can use the STW
identities to arrive at a contribution which is illustrated in Fig.10
where the dotted line brings color to the $(\ell-k-r)$ line as well as
transverse momentum,but it brings no longitudinal momentum and no
momentum dependence at its vertex. This can be interpreted as a color
rotation and transverse momentum convolution of the field of nucleon 1
by the gauge field from nucleon 2 followed by a scattering off the
left-moving nucleon.

Now consider the graphs shown in Fig.11. We again distort the $k_+-$
contour of integration into the upper half plane.  In Fig.11a there is
also the light-cone denominators from the $(k+r)-$line which must be
considered.  The term having a singularity in the upper-half plane is
proportional to ${(k+r)_\lambda\over (k+r)_+-i\epsilon}$.  It is
straightforward to check, using the STW identities, that this term
gives zero because of our choice $\bar{\eta}\cdot\epsilon = 0.$ Thus
in graph(a) there is only the ${k_\alpha\over k_+-i\epsilon}$
singularity of the $k-$line which is evaluated by the STW identities.
The graph shown in Fig.11b has two contributions, one being the
${k_\alpha\over k_+-i\epsilon}$ term from the $k-$ propagator and the
second being a singularity in the nucleon-gluon scattering process
$(P) + (-r-k) \to (P-r) + (-k).$ This second term corresponds to a
single gluon exchange final state interaction of the left- moving
nucleon remnants with nucleon 2 and it cancels with corresponding
two-gluon exchange terms in the amplitude and complex conjugate
amplitude.  The result of the terms in Fig.11 is illustrated in Fig.12
and can be interpreted as a color rotation, and convolution, of the
field from the left-moving nucleon by nucleon 2 followed by a
scattering off nucleon 1.

Now we consider the graphs shown in Fig.13.  We shall evaluate these
graphs by distorting the $k_+-$contour into the {\em lower} half
plane.  Since we now distort in a direction opposite to that natural
for the factor $e^{ik_+(x_2-x_1)_-}$ it is important to check that
there is sufficient convergence in the $k_+-$ plane so that the
exponential factor is not relevant.  For example, in graph(a) one has
$k_+-$ denominators
$[k_+-i\epsilon]^{-1},[(k+r-\ell)_++i\epsilon]^{-1}$ and $[(k-\ell)_++
{(\underline{k}-{\underline \ell})^2\over 2\ell_-}-i\epsilon]^{-1}$
multiplying $k_\alpha^\perp$ and $(k+r-\ell)_\rho^\perp$ as well as a
possible light-cone denominator from the $(\ell-k)-$line.  (Note,
however, that the ${\eta_\nu(k-\ell)_\mu\over (k- \ell)_++i\epsilon}$
term in the $(\ell-k)-$propagator gives a small contribution because
of our choice of polarization,$\bar{\eta}\cdot\epsilon =0.)$ Thus
there is sufficiently strong convergence for the $k_+-$integral unless
(possibly) there is a factor of $k_+$ coming from each of the 3-gluon
vertices.  At the $\Gamma_{\mu\alpha\gamma}$ vertex the only
possibility leads to a factor of $\bar{\eta}_\mu.$ (Factors of
$\bar{\eta}_\alpha$ or $\bar{\eta}_\gamma$ give zero because they
multiply $k_\alpha^\perp$ or $\epsilon_\gamma$ respectively.)  At the
$\Gamma_{\lambda\rho\nu}$ vertex a factor of $\bar{\eta}_\nu$ leads to
an extra denominator $[(k-\ell)_+]^{-1}$ as does a factor of
$\bar{\eta}_\lambda$ since $(k+r-\ell)_\rho^\perp
g_{\rho\nu}\bar{\eta}_\nu = 0.$ Thus there is sufficient convergence
to distort the $k_+-$ contour in the lower half plane where one picks
up only the pole $[(k+r-\ell)_++i\epsilon]^{-1}.$ After picking up
this pole one writes $(k+r-\ell)_\rho^\perp =
(k+r-\ell)_\rho-\eta_\rho(k+r-\ell)_--\bar{\eta}_\rho(k+r-\ell)_+$ and
using $(k+r- \ell)_+=0$ and $(k+r-\ell)_-\approx 0$ one again can use
the STW identities as was done for the graphs of Fig.11.  Here one
obtains a result illustrated in Fig.14 along with a rescattering of
the remnant of the left-moving nucleon which will cancel as discussed
above.  Hence, we may view the graphs of Fig.13 as a color rotation,
and transverse momentum convolution of the field from the left-moving
nucleon by nucleon 1 followed by a scattering off nucleon 2.

Finally, there is the graph shown in Fig.15.  Here we distort the
contour in the upper half plane to pick up the pole at $k_+=0$ with
the factor $k_\alpha$ acting at the 4-gluon vertex after the
$k_+-$integration has been done.

In order to see that the graphs of Figs.9,11,13 and 15 give the same
result as our previous calculation summarized in (58) it is convenient
to evaluate the graphs of Fig.13 in a different way. If we evaluate
these graphs by distorting the $k_+-$ contour into the upper half
$k_+-$plane poles at $k_+=0$ and at $(\ell-k)_+={({\underline
\ell}-\underline{k})^2\over 2\ell_-}$ are encountered.  (There is also the
term ${\eta_\mu(k-\ell)_\nu\over (k-\ell)_+-i\epsilon}$ coming from
the $(\ell-k)-$line propagator which gives an additional singularity,
however this term leads to a small result since the $(k-\ell)_\nu$
term either cancels between the \ $a$\ and \ $b$\ parts of Fig.13 by
STW or the $(k-\ell)_\nu$ eliminates the $(\ell-k-r)-$ propagator
leaving no singularities whatever in the lower half $k_+$plane.)  When
the singularity at $k_+=0$ is taken one has a resulting $k_\alpha$
which when combined with the results given by the graphs in Figs.10,
12 and 15 gives zero from the STW identities.  The singularity at
$(\ell-k)_+={({\underline k}-{\underline \ell})^2\over 2\ell_-}$ gives
on-shell propagation of the $(\ell-k)-$line with a rescattering off
nucleon 2 after production by nucleon 1.  Thus we recover the result
of our calculation in Sec.3.2 when that calculation is restricted to
the interaction of just two nucleons in the nucleus.

However, one can also describe the scattering in terms of the
``classical'' $A_+=0$ light-cone gauge fields of the left-moving
nucleon and of the 2 right-moving nucleons.  In this case the
left-moving nucleon has a classical field color rotated by nucleons 1
and 2 as indicated in Figs.12 and 14.  The right-moving nucleons have
their Weizs\"acker-Williams fields where the field of nucleon 1 is
rotated by that of nucleon 2.  In every case the nucleons with larger
values of $x_-$ rotate the fields coming from smaller values of $x_-$
according to (37) and (38).  The gluon production amplitude is then
given by the three and four gluon interactions of these incoming
``classical'' fields as described by the graphs in Figs.10, 12, 14 and
15.  In Appendix B we shall outline the technical argument for this
result when an arbitrary number of nucleons in the nucleus are
involved in the interaction.

We now put into formulas what we described above for gluon production
in a nucleon- nucleus collision calculated in $A_+=0$ gauge, and where
the nucleon is left-moving and the nucleus is right-moving.  We denote
by $A_\mu^\perp(\underline{k}, k_+)$ the field of the nucleus given by
(37-40) and (42).  The ``classical'' field of the nucleon is denoted
by

$$A_\mu^\prime(\underline{k}, k_-,{\underline b}) =
A^\prime(\underline{k},k_-,{\underline b})\eta_\mu\eqno(68)$$

\noindent with

$$A^\prime({\underline x}, x_+,{\underline b}) = - S({\underline x})
T^aS^{- 1}({\underline x})\delta(x_+) \ell n[\vert{\underline
x}-{\underline b}\vert\mu]{\hat{\rho}}_N^a\eqno(69)$$

\noindent where $S({\underline x})$ is the same as in (38) but with
$x_-=-\infty,$ and where

$$<{\hat{\rho}}_N^a{\hat{\rho}}_N^{a^\prime}> =
{\delta_{aa^\prime}\over N_c^2-1} Q^2\ {\partial\over \partial Q^2}
xG(x,Q^2)\eqno(70)$$

\noindent indicates an average over internal nucleon structure.  We
have used current conservation at the nucleon source to eliminate the
light-cone gauge denominator in $A_\mu^\prime$ much as was done in the
discussion following (53) and which led to (54), but now with the
roles of + components and --- components exchanged.  In order to write
a compact formula for ${d\sigma\over d^2\ell dy}$ it is convenient to
define a total gluon field $A_\mu^{tot}(x)$ as

$$A_\mu^{tot}(x) = A_\mu^\perp(x) + A_\mu^\prime(x,{\underline b}) +
A_\mu^{free}(x)\eqno(71)$$

\noindent where $A^\perp$ and $A^\prime$ are the fields given in (37)
and (68), respectively, while $A_\mu^{free}(x)$ is a free quantized
gluon field normalized according to

$$(0\vert A_\mu^{free}(x)\vert\ell\lambda a) =
{\epsilon_\mu^{(\lambda)}(\ell) e^{i\ell\cdot x}\over \sqrt{(2\pi)^3
2\omega_\ell}} T^a\eqno(72)$$

\noindent with $\vert\ell\lambda a)$ being a single gluon state having
momentum $\ell,$ polarization $\lambda$ and color $a.$ $A^{tot} $
depends on the impact parameter of the collision through $A^\prime.$
Then the gluon production cross section is

$${d\sigma\over d^2\ell dy} = \omega_\ell \int d^2b
<\sum_{\lambda=1}^2\ \sum_{a=1}^{N_c^2-1}(0\vert S\vert\ell\lambda
a)(\ell \lambda a\vert S\vert\ 0)>\eqno(73)$$

\noindent with

$$S = {-1\over 2} \int d^4x\ Tr F_{\mu\nu}(x)F_{\mu\nu}(x)\eqno(74)$$

\noindent and where

$$F_{\mu\nu}=\partial_\mu A_\nu^{tot} - \partial_\nu A_\mu^{tot} -
ig[A_\mu^{tot}, A_\nu^{tot}].\eqno(75)$$

\noindent The $< >$ in (73) indicate an average over nuclear structure
as indicated in (39) as well as an average over nucleon structure as
given by (70).  In evaluating (73) one assumes that the field
$A_\mu^{free}$ is much smaller than $A_\mu^\prime$ and $A_\mu^\perp.$
Therefore $A_\mu^{free}$ is included only once in the 3 and 4 gluon
vertices in S.  At the same time each of the other two fields has to
be included at least once to allow the production of the gluon.  Eq.73
is a compact, and elegant, formula for gluon production in
nucleon-nucleus collisions.

\clearpage
\noindent{\Large\bf Appendix A}
\vskip 20pt
\noindent In this appendix we shall show how the exponential factors
in (57) come about.  We consider the case where $\tau_1<0$ and
$\tau_2> 0.$ We consider the last, the latest in time, interaction
with a nucleon in the nucleus.  In the amplitude the interaction is
with a quark-gluon system with the quark at transverse coordinate
$\b{0}$ and the gluon at ${\underline x}_1.$ In the complex conjugate
amplitude the interaction is only with the quark, again at transverse
coordinate $\b{0}.$ The possible interaction terms, both real and
virtual, are shown in Fig.16.  The factors associated with each graph
are listed below as

$$(a)	=	- C_F/2\  \tilde{V}(\b{0})/N_c$$
$$(b)	=   \ \ N_c/2\  \tilde{V}({\underline x}_1)/N_c$$
$$(c)	=	-1/2N_c\  \tilde{V}(\b{0})/N_c\eqno(A.1)$$
$$(d)	=	-	N_c/2\  \tilde{V}(\b{0})/N_c$$
$$(e)	=  \ \  N_c/2\  \tilde{V}({\underline x}_1)/N_c$$
$$(f) = - C_F/2\  \tilde{V}(\b{0})/N_c$$

\noindent where $\tilde{V}$ is given in (17).  Adding all the terms
one finds

$$(a) + (b) \cdot \cdot \cdot + (f) =\ -
[\tilde{V}(0)-\tilde{V}({\underline x}_1)]= - {1\over 4}
x_1^2\tilde{v}({\underline x}_1).\eqno(A.2)$$

\noindent Multiplying the right-hand side of (A.2) by $\rho\sigma =
1/\lambda$ to take account of the density of nucleons and the cross
section we obtain the same type of factor as appears on the right-hand
side of (14).  Our convention is to use $\sigma,$ and $\lambda,$ for
gluon scattering on nucleons.  The factor of $1/N_c$ on the right-hand
side of the terms in (A.1) converts the result from gluons to the
appropriate partons involved in the interaction. \\~\\~\\

\noindent {\Large\bf Appendix B}
\vskip 20pt In this appendix we outline an argument for the absence of
final state interactions in gluon production in nucleon-nucleus
collisions.  The major barrier in generalizing the discussion given in
Sec.3.3, for the absence of final state interactions when a nucleon
scatters on two nucleons, is the presence of light-cone gauge
denominators of the type ${1\over (k-\ell)_++i\epsilon},$ along the
gluon line carrying the large minus-component of the momentum, which
hinder the distortion of $k_+-$ contours into the lower half plane.
In Sec.3.3, for the graphs in Fig.13 this term was eliminated by our
choice of polarization vector, $\bar{\eta}\cdot \epsilon = 0$ for the
produced gluon.  In the general case the troublesome graphs are of the
type shown in Fig.17.  The graph in Fig.17a is such that cutting the
$(\ell-k)$-line separates the overall graph into two parts while the
graph in Fig.17b is an example of a graph where cutting the
$(\ell-k)$-line does not separate the graph into two parts.  There is
not any particular ordering of the $x_--$ positions of the nucleons
belonging to \ $A$\ and \ $B.$\ That is, some nucleons in \ $A$\ may
have larger $x_--$ values than some of the nucleons in \ $B.$\ We
begin with the graphs of the type shown in Fig.17a.  For our purposes
the dangerous term is the ${\eta_{\nu}(k- \ell)_\mu\over
(k-\ell)_++i\epsilon}$ part of the $(\ell-k)-$ propagator.  However,
once all hookings of the $(\ell-k)-$line in \ $B$\ are included, at a
given order of the coupling and with a given set of nucleons, the net
result must be zero from the STW identities.

Now turn to graphs of the type shown in Fig.17b where one or more
lines, in addition to the ($\ell-k)-$line connect\ $A$\ and\ $B.$\ A
particular example of such a graph is shown in Fig.18, where nucleon\
1\ and nucleon\ 3\ have inelastic reactions while nucleon 2 has an
elastic reaction.  The potentially dangerous problem with graphs of
this type is that the application of the STW identities does not
directly give zero since the $k_1$ line carries color.  Indeed,
applying the STW identities for the ${-\eta_\nu(k-\ell)_\mu\over
(k-\ell)_++i\epsilon}$ term in the $(\ell-k)-$ propagator leads to a
term like that indicated in Fig.19 where the dashed line carries the
propagator factors ${-i\over (k-\ell)^2+i\epsilon}{-\eta_\nu\over
(k-\ell)_++i\epsilon}$ and inserts a momentum $\ell- k$ on the
$k_1-$line as well as rotating the color factors of the $k_1-$line.
However, this term is small because the scattering with nucleon 2 is
the same as the high energy scattering gluon $(\ell-k-k_1) +$ nucleon
$\to$ gluon $(\ell-k_2) +$ nucleon and the high energy limit of such a
scattering is zero in our approximation of no evolution within nucleon
2.

Thus, the gluon line carrying the large minus component of the
momentum $\ell,$ the hard gluon line, will never give singularities as
one distorts $k_+-$momenta into the lower half plane.  This is all
that is needed to arrive at our main result, described in Sec.3.3 and
summarized in (73)-(75).  To see this, begin with the nucleon having
the largest values of $x_-.$ A gluon,\ $g,$\ from this nucleon either
attaches directly to the hard gluon line or to a gluon coming from
another nucleon or, perhaps, directly to another nucleon. If the
gluon, $g$,\ does not attach directly to the hard gluon one is in the
case considered in Ref.11 where it was shown that\ $g$\ rotates the
fields coming from nucleons with smaller values of $x_-.$ If the
gluon,\ $g,$\ attaches to the left-moving nucleon, or to the hard
gluon before the hard gluon has any previous interactions, one
distorts the $k_+$ of\ $g$\ into the upper half plane and uses the STW
identities.  If \ $g$\ attaches to the hard gluon after the hard gluon
has interaction with other nucleons one routes the momentum of \ $g$\
through the hard gluon and back into a nucleon at the first available
interaction, as illustrated for example in Fig.18.  In this case one
distorts the $k_+-$contour in the lower half plane and uses the STW
identities not for \ $g$\ but for the gluons carrying the momentum\
$k$\ returning from the hard gluon.  This set of operators leads to
the fields $A_\mu^\perp$ and $A_\mu^\prime$ in (71) and a cross
section given by (73).

\begin{figure}
\begin{center}
\leavevmode
\hbox{ \epsffile{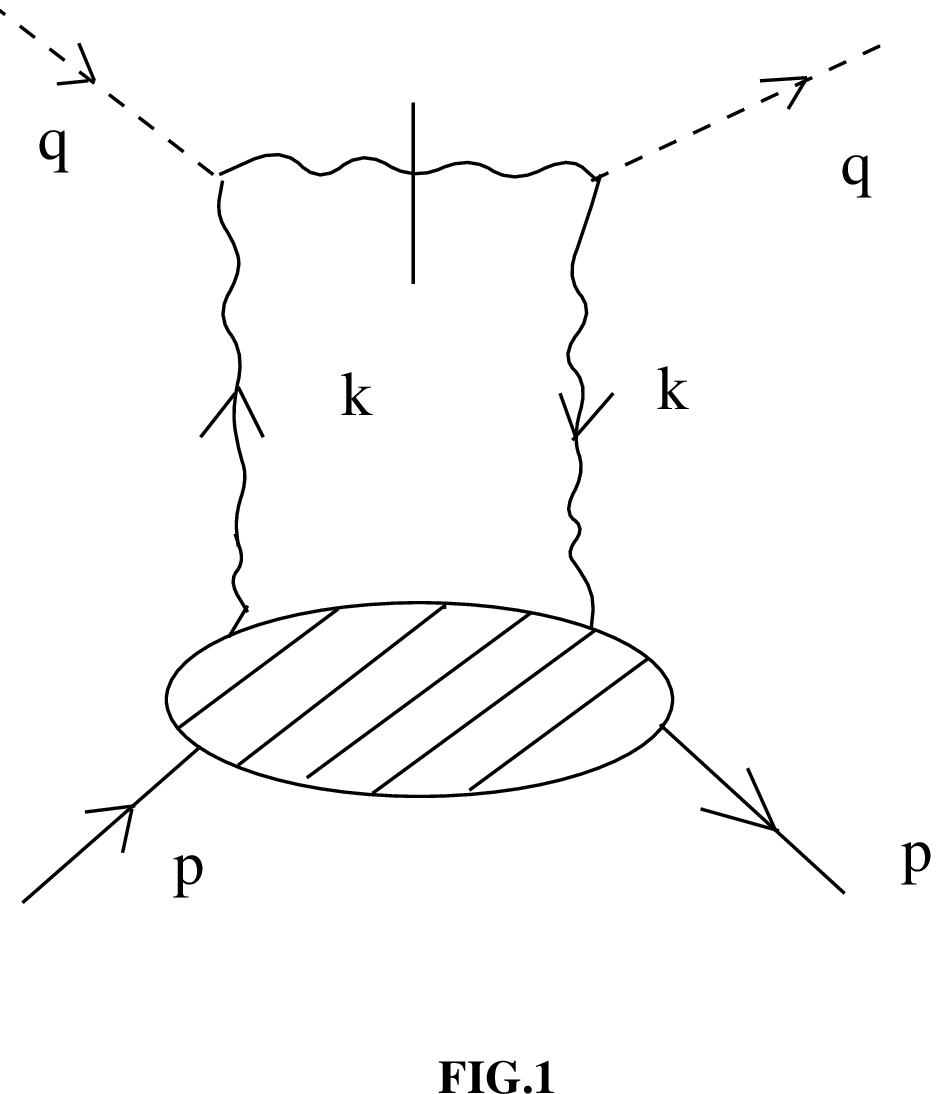}}
\end{center}
\end{figure}

\pagebreak

\begin{figure}
\begin{center}
\leavevmode
\hbox{ \epsffile{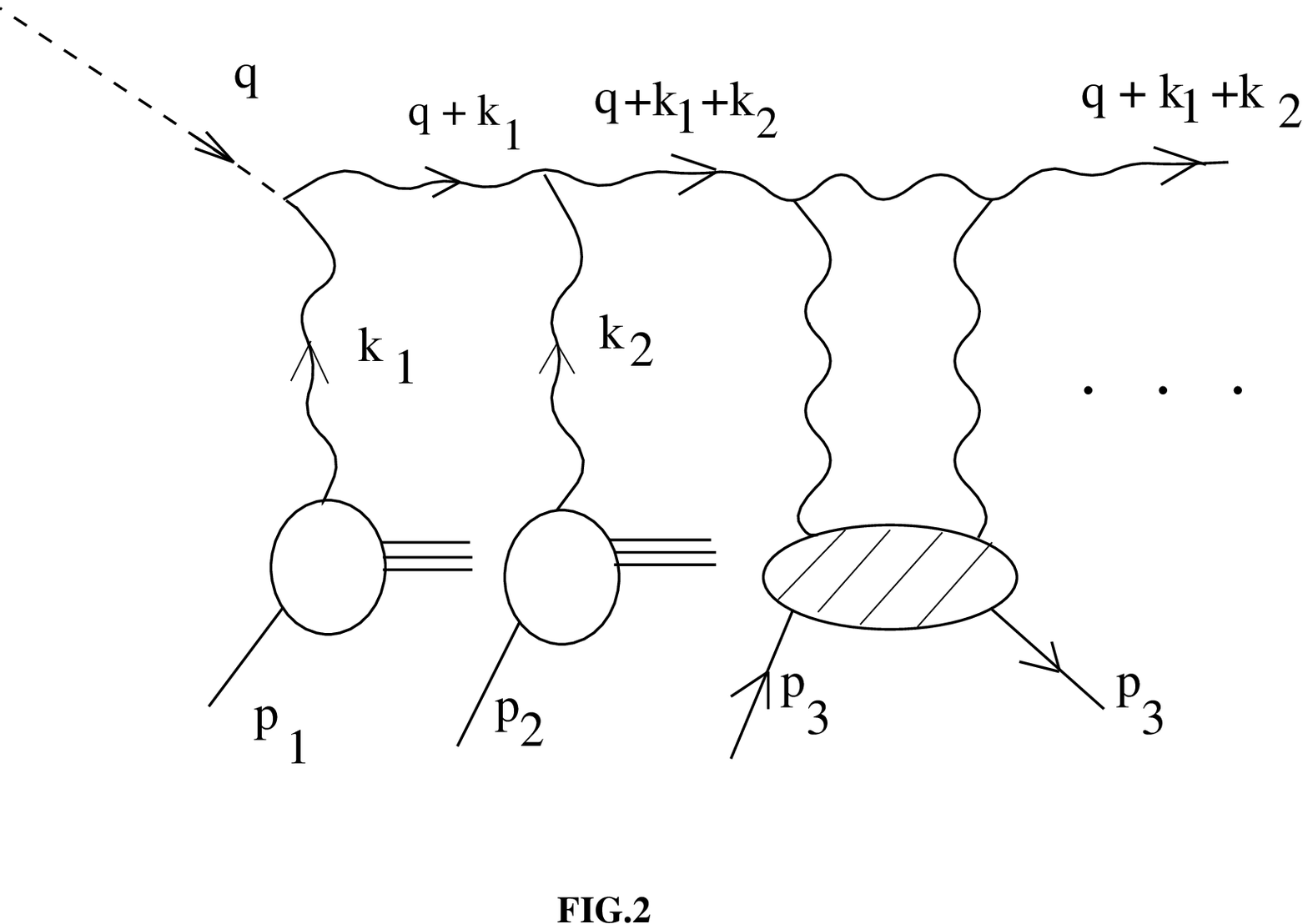}}
\end{center}
\end{figure}

\pagebreak

\begin{figure}
\begin{center}
\leavevmode
\hbox{ \epsffile{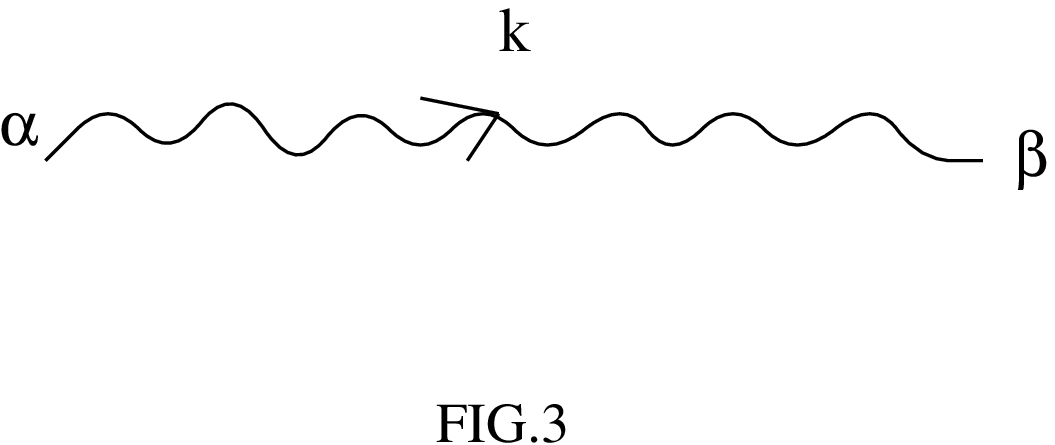}}
\end{center}
\end{figure}

\pagebreak

\begin{figure}
\begin{center}
\leavevmode
\hbox{ \epsffile{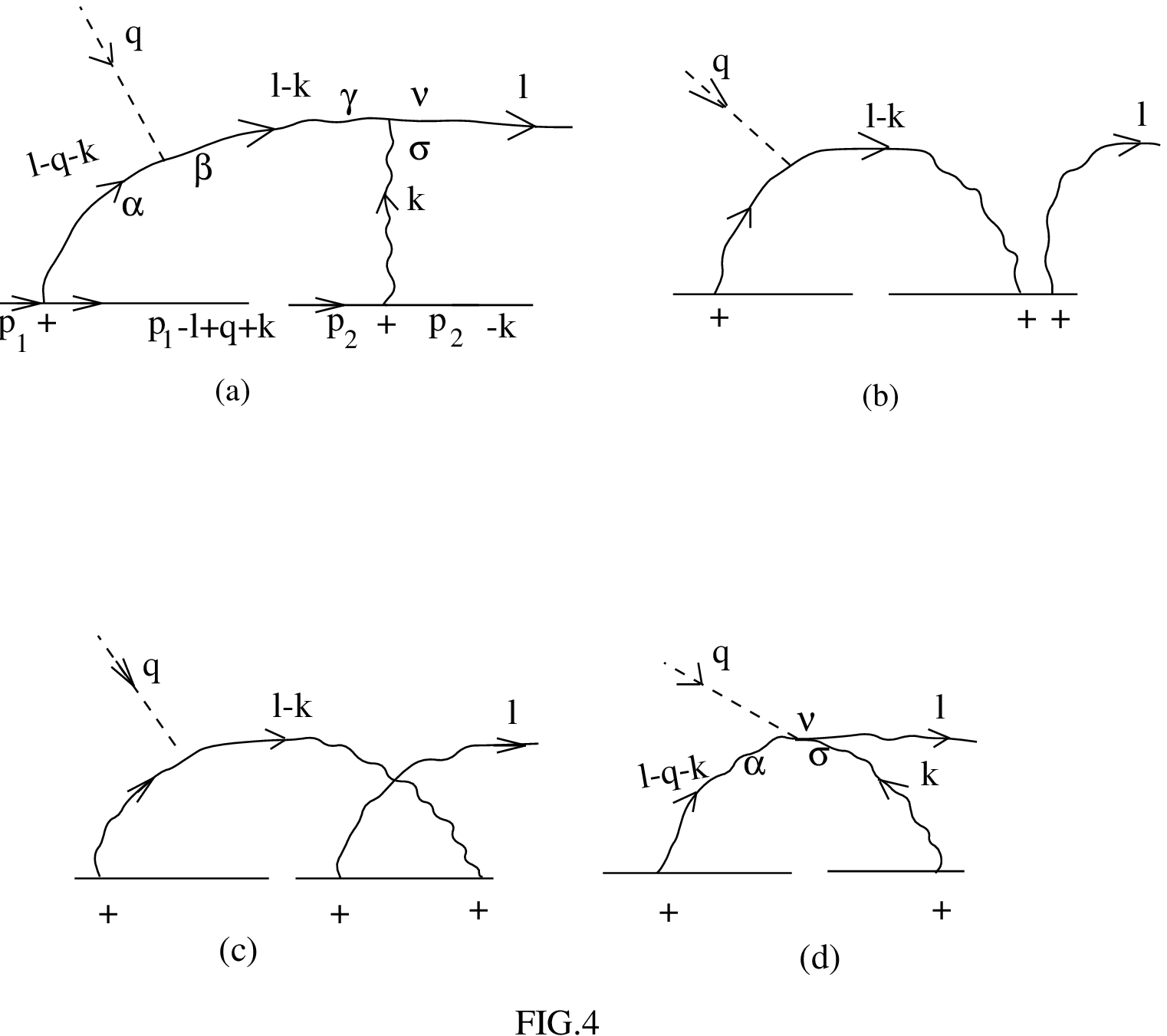}}
\end{center}
\end{figure}

\pagebreak

\begin{figure}
\begin{center}
\leavevmode
\hbox{ \epsffile{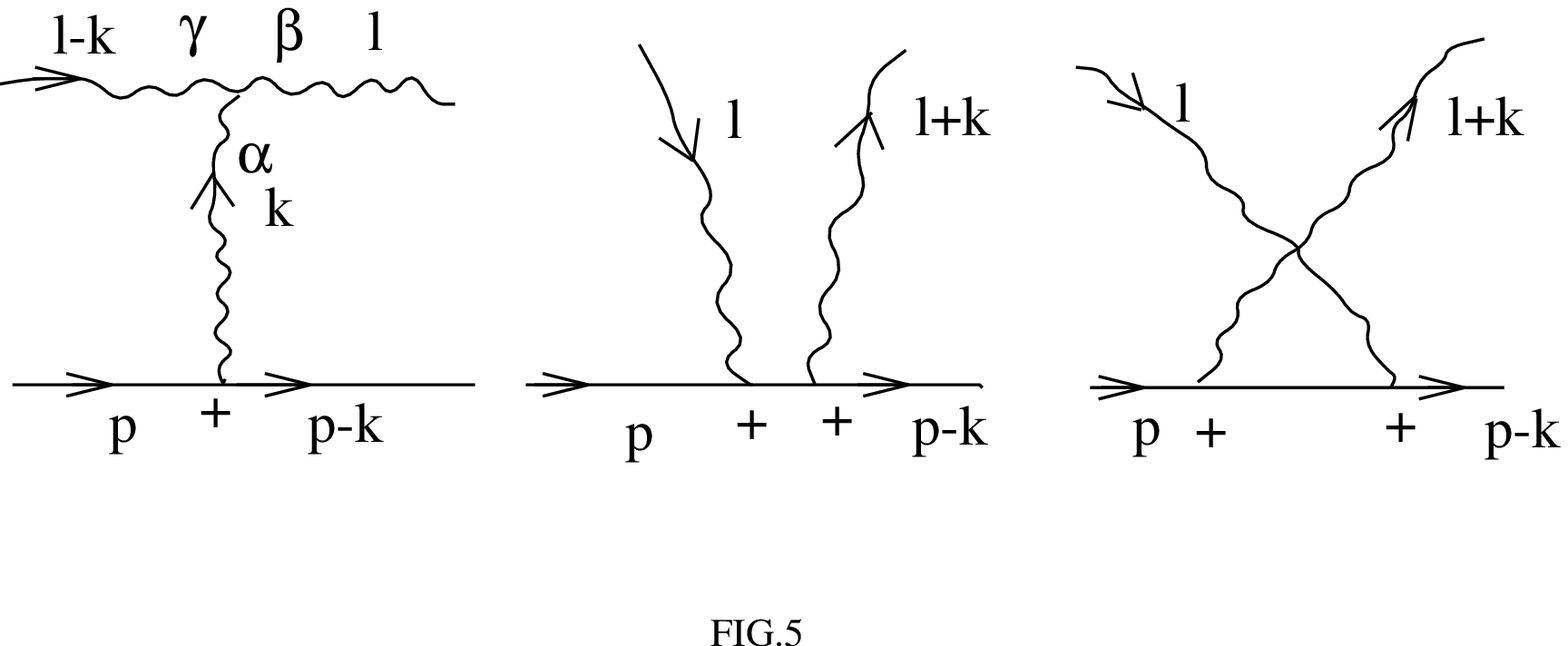}}
\end{center}
\end{figure}

\pagebreak

\begin{figure}
\begin{center}
\leavevmode
\hbox{ \epsffile{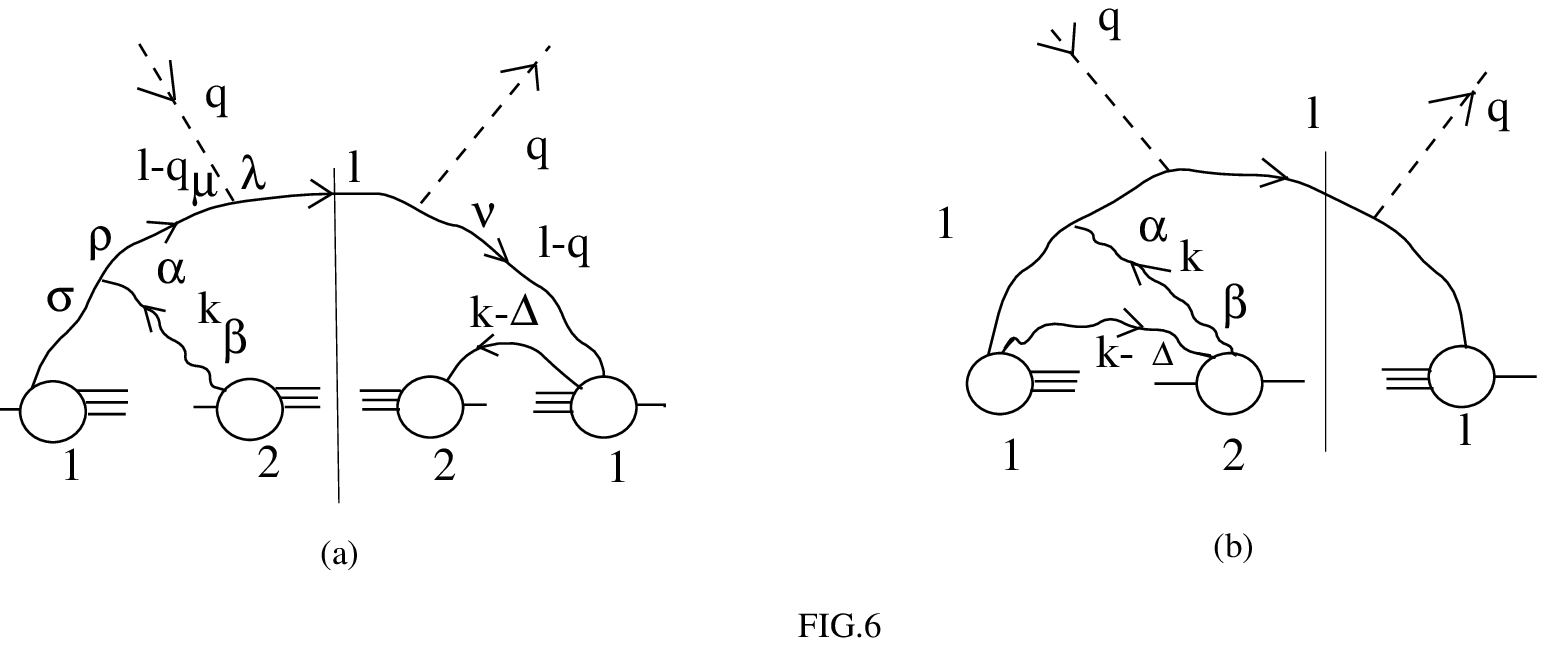}}
\end{center}
\end{figure}

\pagebreak

\begin{figure}
\begin{center}
\leavevmode
\hbox{ \epsffile{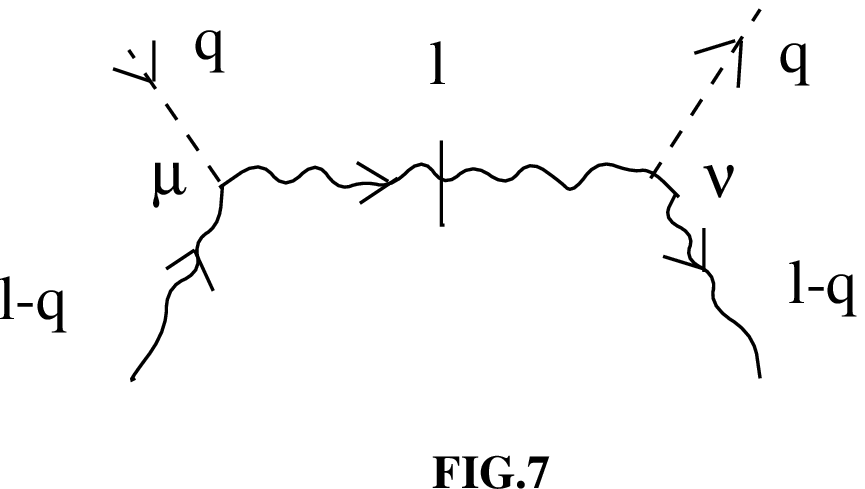}}
\end{center}
\end{figure}

\pagebreak

\begin{figure}
\begin{center}
\leavevmode
\hbox{ \epsffile{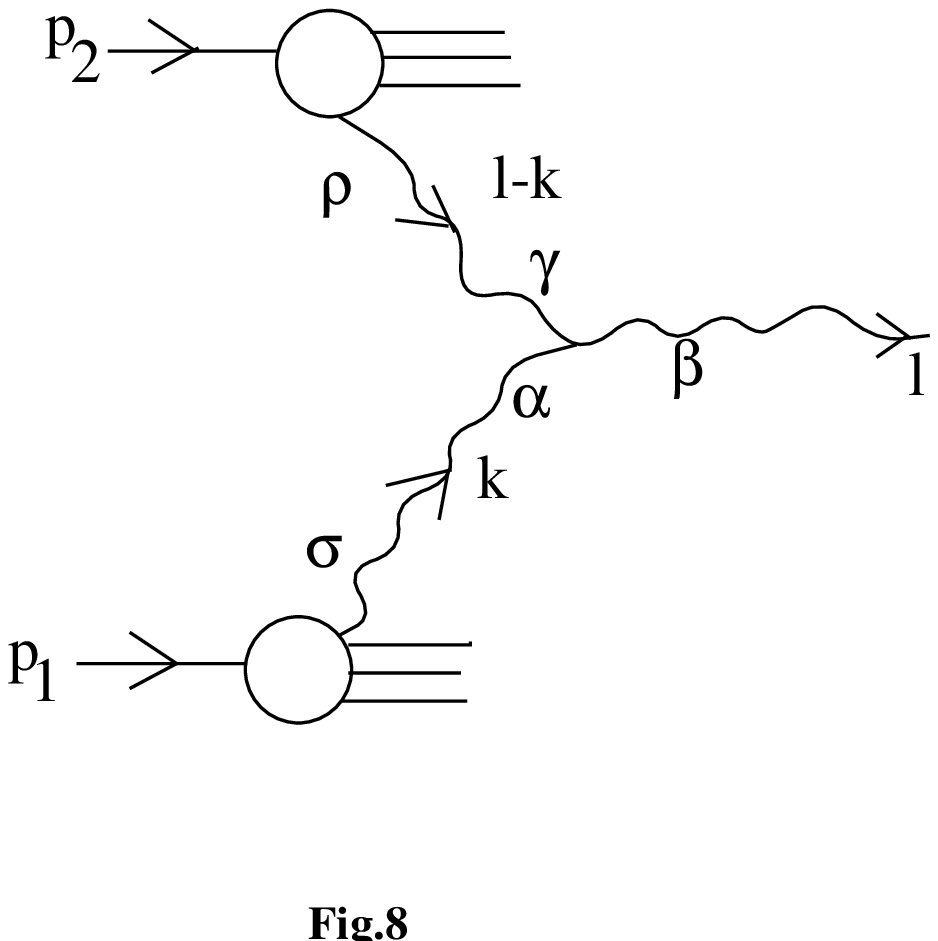}}
\end{center}
\end{figure}

\pagebreak

\begin{figure}
\begin{center}
\leavevmode
\hbox{ \epsffile{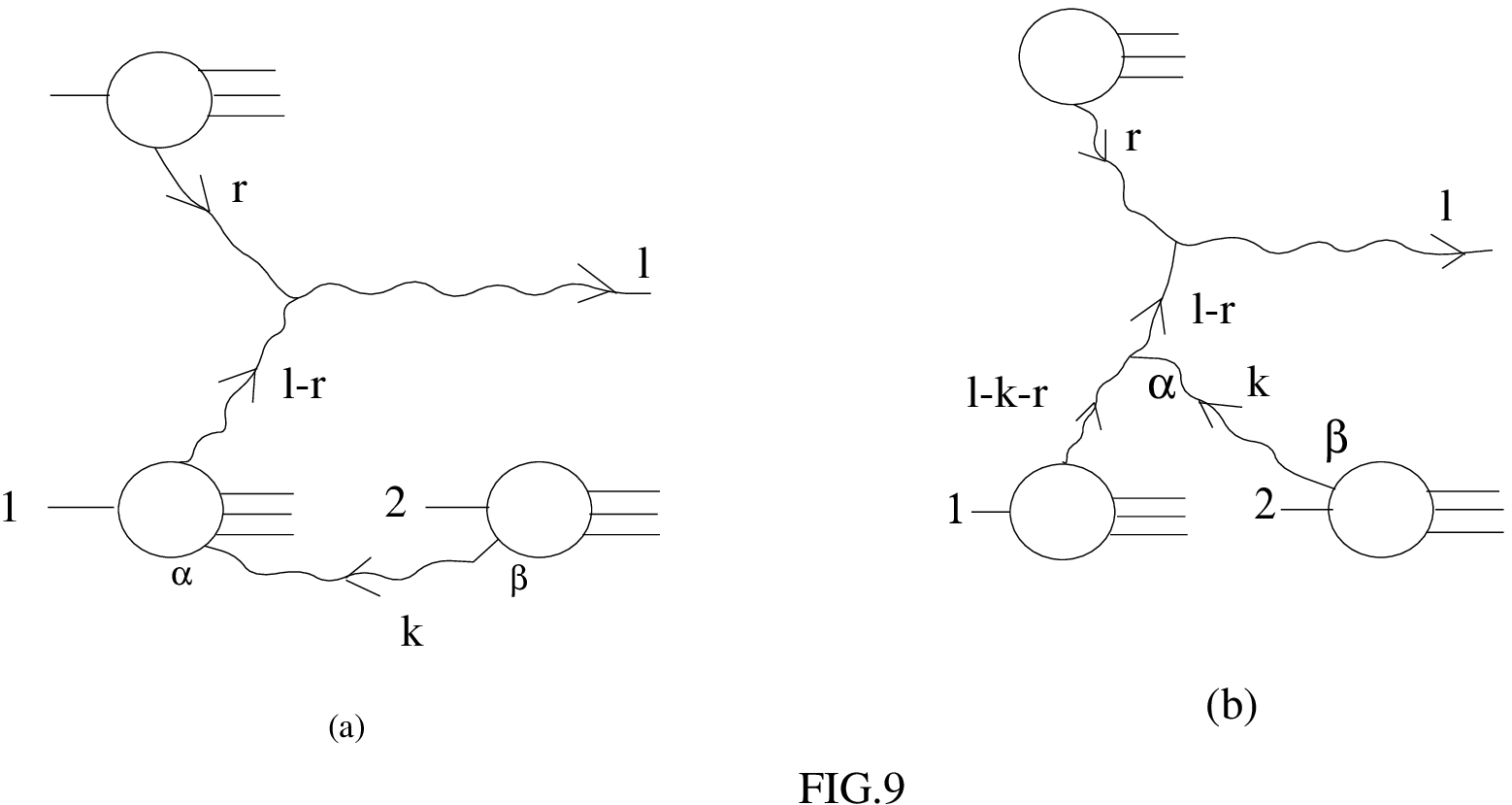}}
\end{center}
\end{figure}

\pagebreak

\begin{figure}
\begin{center}
\leavevmode
\hbox{ \epsffile{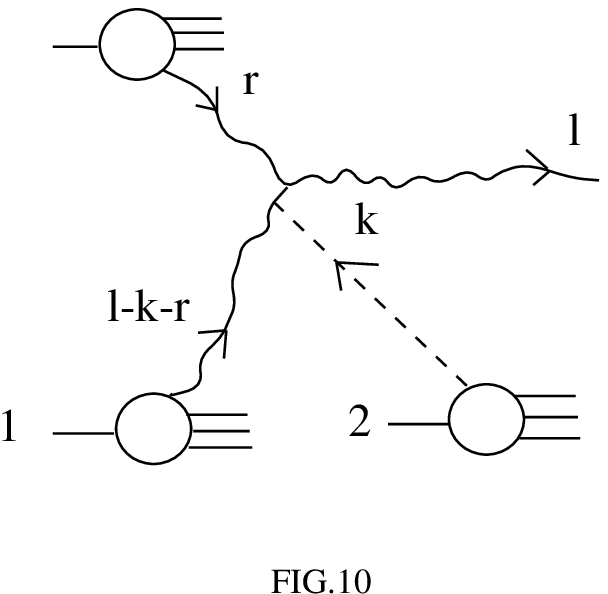}}
\end{center}
\end{figure}

\pagebreak

\begin{figure}
\begin{center}
\leavevmode
\hbox{ \epsffile{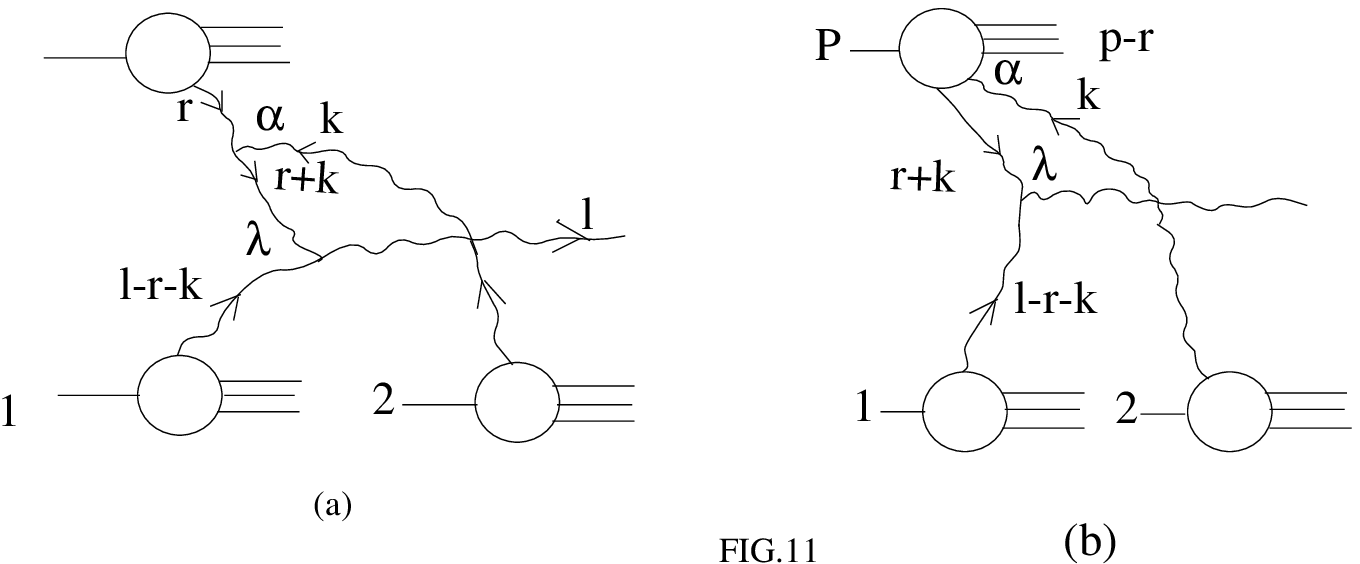}}
\end{center}
\end{figure}

\pagebreak

\begin{figure}
\begin{center}
\leavevmode
\hbox{ \epsffile{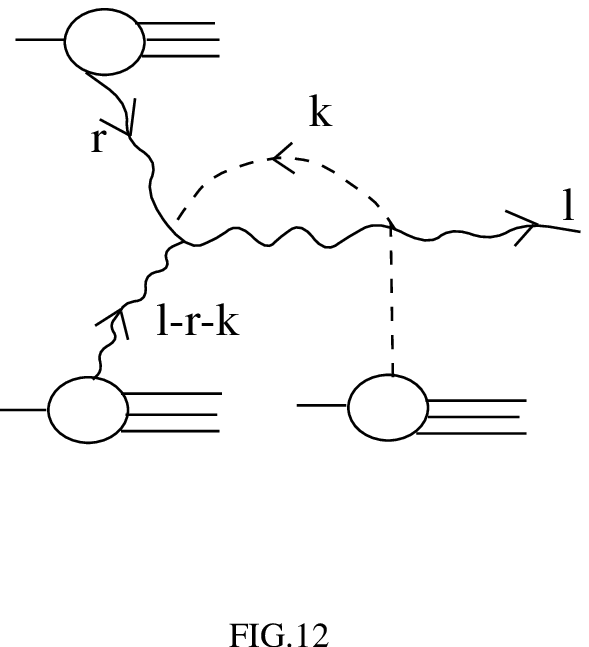}}
\end{center}
\end{figure}

\pagebreak

\begin{figure}
\begin{center}
\leavevmode
\hbox{ \epsffile{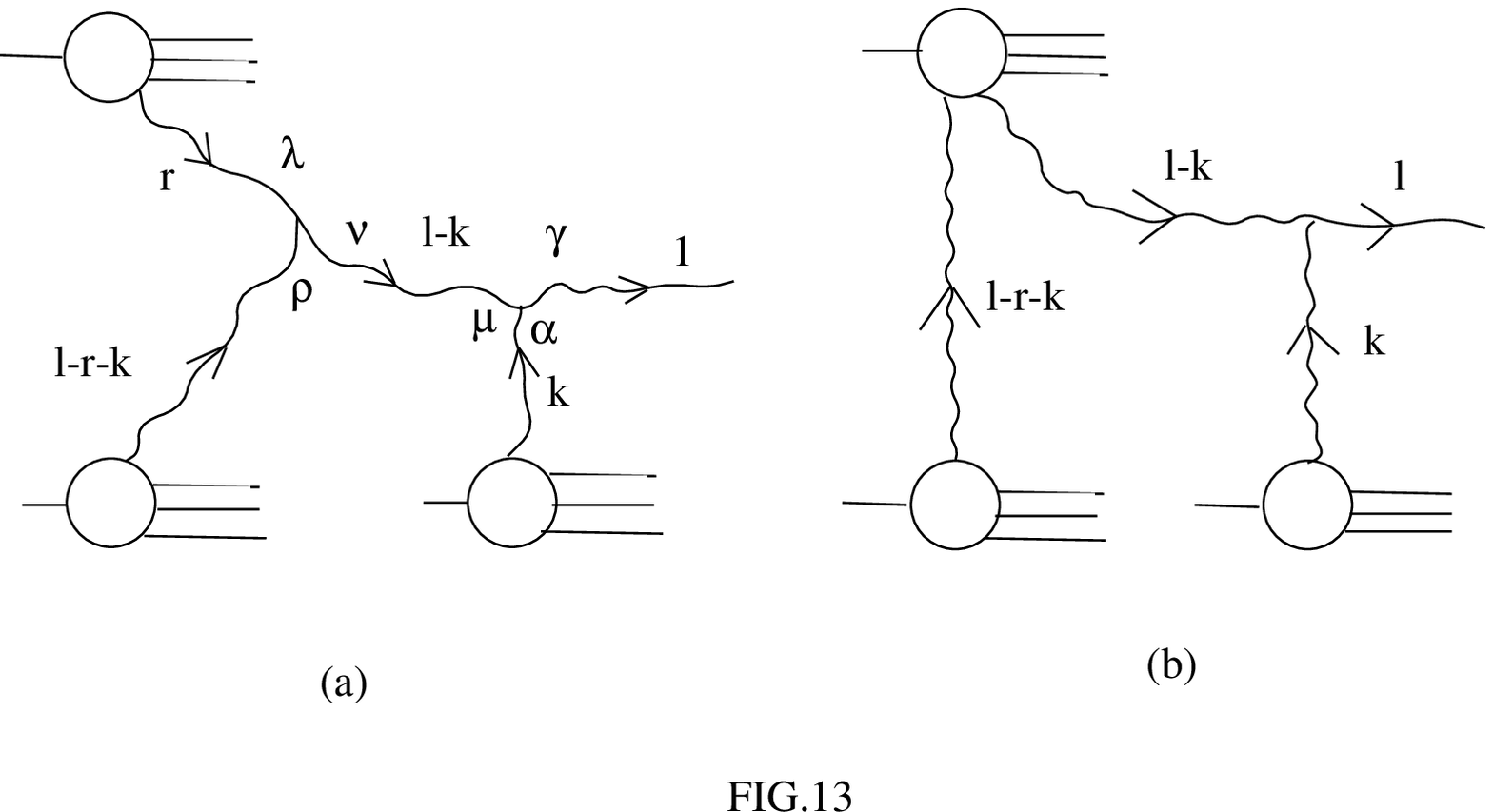}}
\end{center}
\end{figure}

\pagebreak

\begin{figure}
\begin{center}
\leavevmode
\hbox{ \epsffile{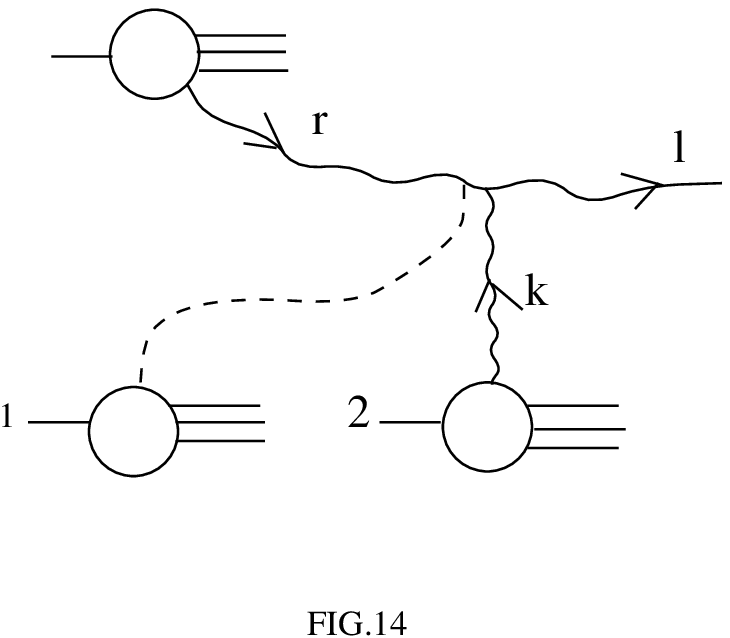}}
\end{center}
\end{figure}

\pagebreak

\begin{figure}
\begin{center}
\leavevmode
\hbox{ \epsffile{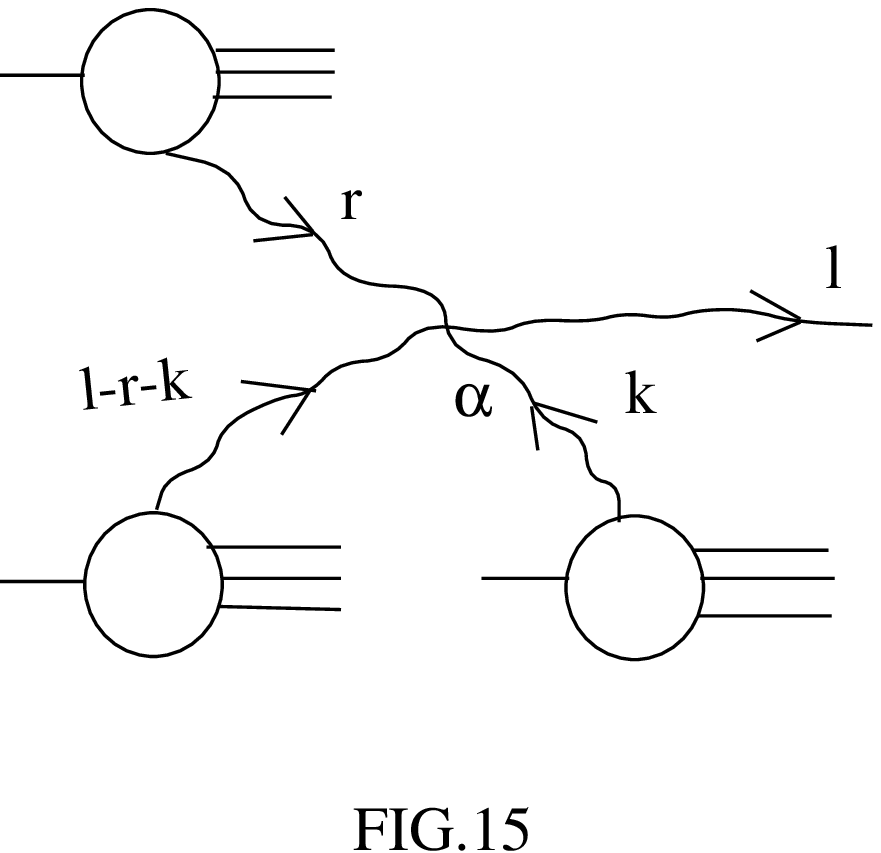}}
\end{center}
\end{figure}

\pagebreak

\begin{figure}
\begin{center}
\leavevmode
\hbox{ \epsffile{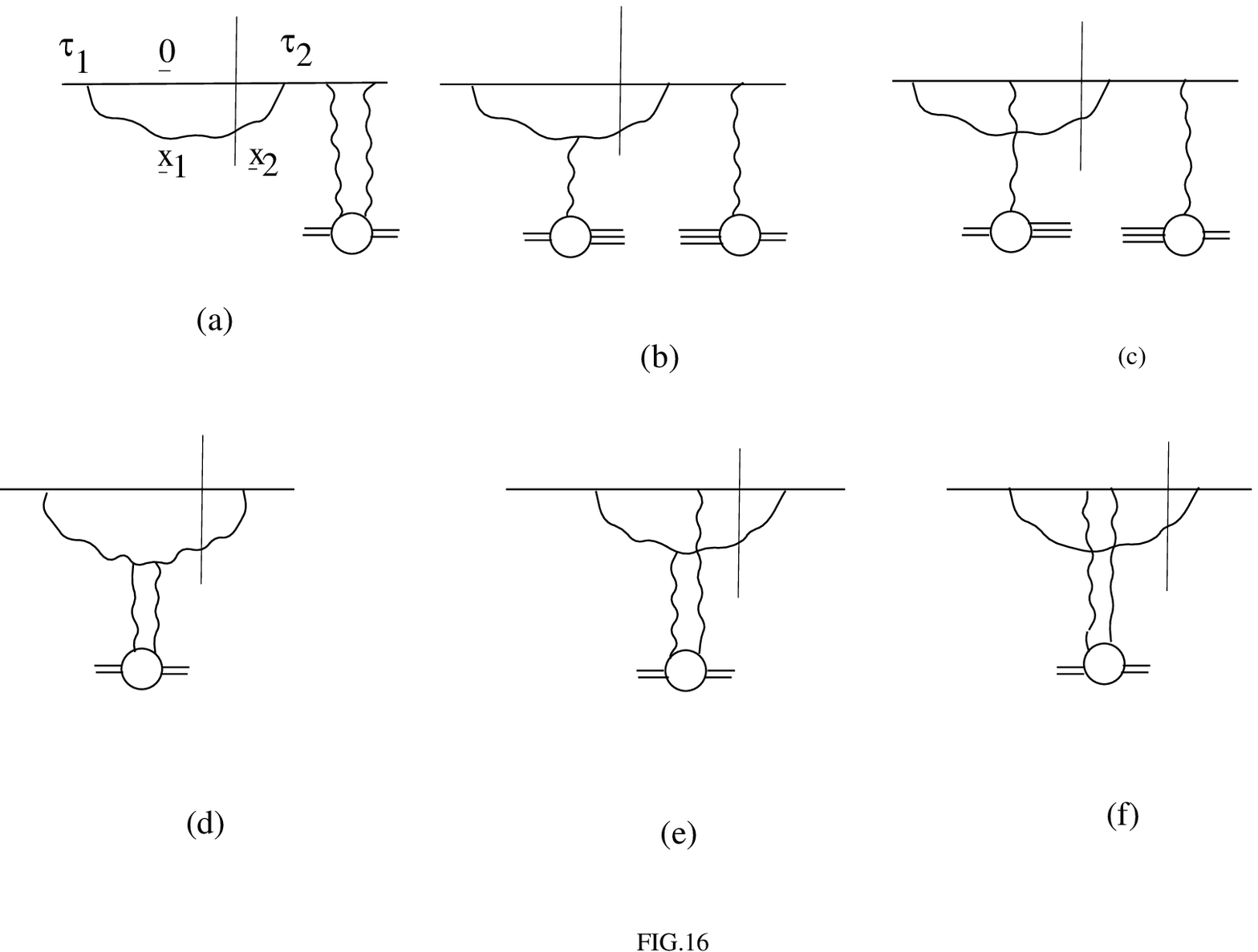}}
\end{center}
\end{figure}

\pagebreak

\begin{figure}
\begin{center}
\leavevmode
\hbox{ \epsffile{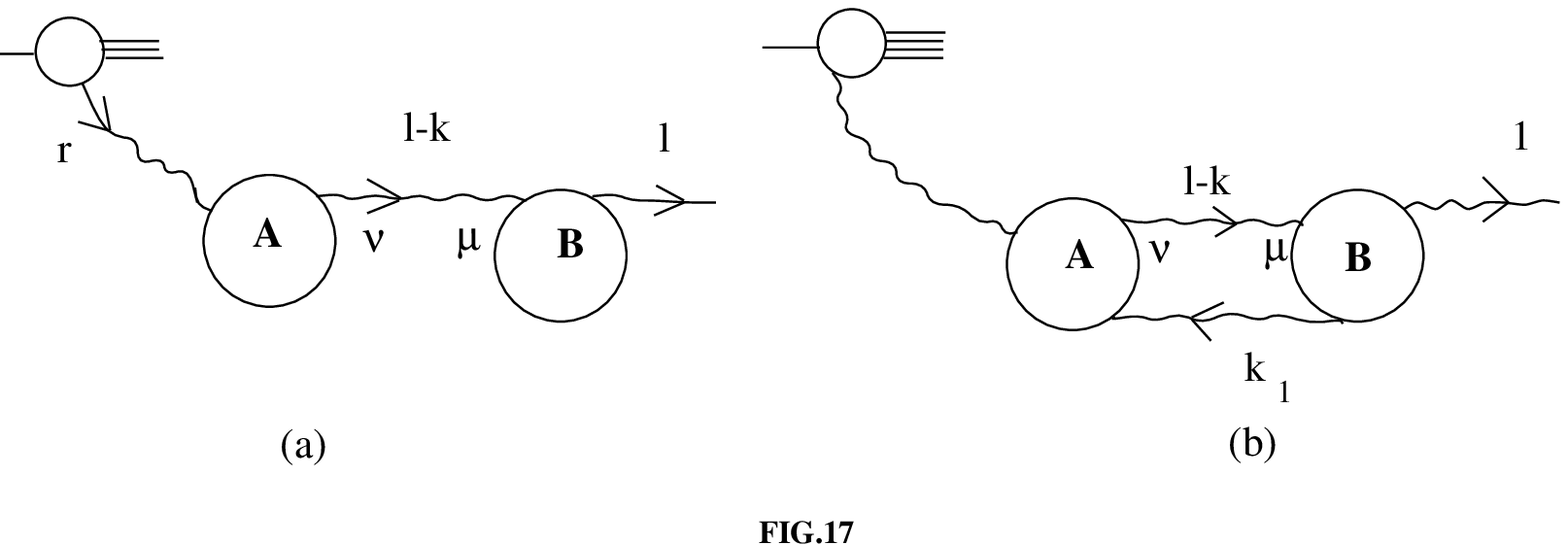}}
\end{center}
\end{figure}

\pagebreak

\begin{figure}
\begin{center}
\leavevmode
\hbox{ \epsffile{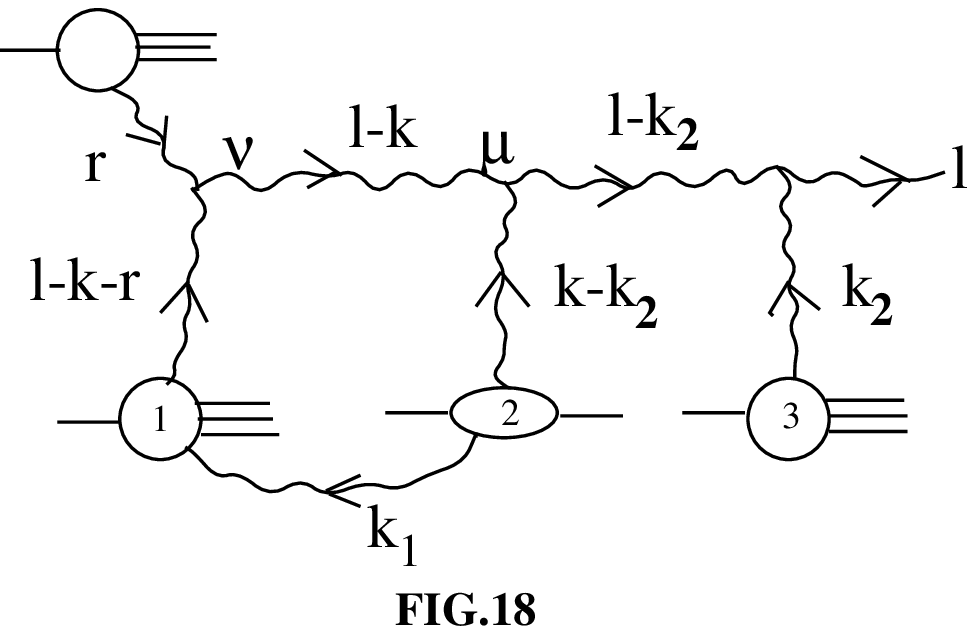}}
\end{center}
\end{figure}

\pagebreak

\begin{figure}
\begin{center}
\leavevmode
\hbox{ \epsffile{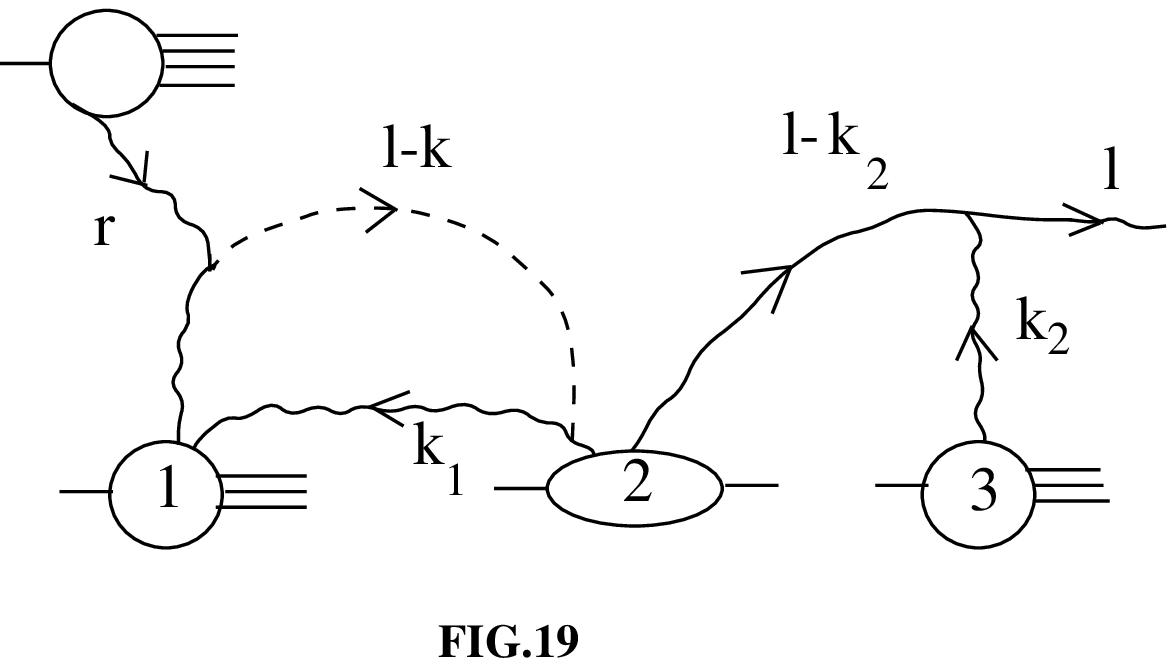}}
\end{center}
\end{figure}

\pagebreak

\end{document}